\documentclass[structabstract]{aa}
\usepackage{graphicx}
\usepackage{txfonts}
\usepackage{natbib}
\bibpunct{(}{)}{;}{a}{}{,}

\usepackage{longtable}
\usepackage{amssymb}
\usepackage[latin1]{inputenc}
\usepackage{lscape}

\def\kms{km~s$^{-1}\,$}
\def\18182{IRAS~18182--1433}
\def\G16{G16.59--0.05}
\def\Jyb{Jy~beam$^{-1}$}

\titlerunning{Studying Massive Star Formation through maser VLBI}
\begin{document}
   \title{VLBI study of maser kinematics in high-mass SFRs. \textrm{I}. G16.59--0.05}


  \author{ A. Sanna \inst{1,} \inst{2}
\and L. Moscadelli \inst{3} \and R. Cesaroni \inst{3} \and A. Tarchi \inst{2} \and R. S. Furuya \inst{4}  \and C. Goddi \inst{5,} \inst{6}}
   \offprints{A. Sanna, \email{asanna@ca.astro.it}}

   \institute{Dipartimento di Fisica, Università degli Studi di Cagliari, S.P. Monserrato-Sestu km 0.7, I-09042 Cagliari, Italy
   \and INAF, Osservatorio Astronomico di Cagliari, Loc. Poggio dei Pini, Str. 54, 09012 Capoterra (CA), Italy
   \and INAF, Osservatorio Astrofisico di Arcetri, Largo E. Fermi 5, 50125 Firenze, Italy
   \and Subaru Telescope, National Astronomical Observatory of Japan, 650 North A'ohoku Place, Hilo, HI 96720
   \and European Southern Observatory, Karl-Schwarzschild-Strasse 2, D-85748 Garching bei M$\ddot{\rm u}$nchen, Germany
   \and Harvard-Smithsonian Center for Astrophysics, 60 Garden Street, Cambridge, MA 02138, USA}


  \abstract
   {}
   {To study the high-mass star-forming process, we have started a large project to unveil the gas kinematics close to young stellar objects (YSOs) through the Very Long Baseline Interferometry (VLBI) of maser associations. By comparing the high spatial resolution maser data, tracing the inner kinematics of the (proto)stellar cocoon, with interferometric thermal data, tracing the large-scale environment of the hot molecular core (HMC) harbouring the (proto)stars, we can investigate the nature and identify the sources of large-scale motions. The present paper focuses on the high-mass star-forming region \G16.}
   {Using the VLBA and the EVN arrays, we conducted phase-referenced observations of the three most powerful maser species in \G16: H$_2$O at 22.2~GHz (4 epochs), CH$_3$OH at 6.7~GHz (3 epochs), and OH at 1.665~GHz (1 epoch). In addition, we performed high-resolution ($\geq0\farcs1$), high-sensitivity ($<0.1$~mJy) VLA observations of the radio continuum emission from the star-forming region at 1.3 and 3.6~cm.}
   {This is the first work to report accurate measurements of the \emph{relative} proper motions of the 6.7~GHz CH$_3$OH masers.
The different spatial and 3-D velocity distribution clearly indicate that the 22~GHz water and 6.7~GHz methanol masers are tracing different kinematic environments.  The bipolar distribution of 6.7~GHz maser l.o.s. velocities and the regular pattern of observed proper motions suggest that these masers are tracing rotation around a central mass of about 35~M$_{\odot}$. The flattened spatial distribution of the 6.7~GHz masers, oriented  NW--SE, suggests that they can originate in a disk/toroid rotating around the massive YSO which drives the $^{12}$CO~(2--1) outflow, oriented NE--SW, observed on arcsec scale.
The extended, radio continuum source observed close to the 6.7~GHz masers could be excited by a wide-angle wind emitted from the YSO associated
with the methanol masers, and such a wind is proven to be sufficiently energetic to drive the NE--SW $^{12}$CO~(2--1) outflow.
The H$_2$O masers distribute across a region offset about 0\farcs5 to the NW of the CH$_3$OH masers, in the same area where emission of high-density molecular tracers, typical of HMCs, was detected. We postulate that a distinct YSO, possibly in an earlier evolutionary phase than that exciting the methanol masers, is responsible for the excitation of the water masers and the HMC molecular lines.}
   {}


   \maketitle
%

\section{Introduction}

The process of massive star formation is based on a fine tuning between the gravitational force, that sets in the collapse of a Jeans-critical cloud, against the different types of pressures, thermal, magnetic, turbulence, and radiation, that regulate the time scale of the accretion (for recent reviews see, e.g., \citealt{Zinnecker2007,Beuther2007}).
To characterize the relationship between accretion-ejection phenomena it is essential to resolve the dynamical structures close to the central engine.
Maser emission from several molecular lines is an useful signpost of the hot, dusty environment (i.e. hot molecular core, HMC) where massive star formation takes place,
which, because of the extremely high dust extinction, could not be investigated with optical and near-infrared (NIR) diagnostics. Maser emission is concentrated in narrow (usually $\lesssim 1$~\kms\ broad), strong (up to 10$^6$~Jy) lines and arises from compact emission centers (``maser spots'') which have typical sizes of a few AU (e.g., \citealt{Minier2002,Moscadelli2003}).
The strong brightness of the maser emission allows us to use the Very Long Baseline Interferometry (VLBI) technique to determine the position of the maser spots with milli-arcsec accuracy, and comparing positions at different VLBI epochs gives accurate measurement of the spot proper motions. Combining proper motions with the maser line-of-sight (l.o.s) velocities (derived via Doppler shift of the observed maser frequency),  one can reconstruct the full 3-D kinematics of the masing gas.
With this in mind, we have started a large project whose final aim is drawing a comprehensive picture of the dynamical processes in high-mass star-forming regions (HMSFRs), by relating the pc-scale phenomena, traced by thermal continuum and molecular lines, with the AU-scale kinematics of the gas around young stellar objects (YSOs), traced by maser emissions.

In particular, observing targets where different molecular maser species appear to be spatially associated, we can hope to better sample the (proto)stellar environment (e.g., \citealt{Goddi2007, Moscadelli2007}). As a first step, we focus on the three most powerful maser transitions, that of water (H$_2$O) at 22.2~GHz, methanol (CH$_3$OH) at 6.7~GHz, and hydroxyl (OH) at 1.665~GHz. H$_2$O masers typically trace fast shocks (of several tens of \kms) excited by either collimated jets found at the base of molecular outflows, or wide-angle winds emitted from the YSOs (e.g., \citealt{Gwinn1992, Torrelles2003, Moscadelli2007}). OH masers are likely associated with the slow expanding (typically a few \kms) ionization front of H~\textsc{ii} regions (e.g., \citealt{Fish2007}). The origin of CH$_3$OH masers, instead, is currently a matter of debate. To date,
single-epoch VLBI observations have provided accurate spatial and l.o.s. velocity distribution of the 6.7~GHz masers towards a few tens of sources. To account for the
ordered gradient of l.o.s. velocities observed in some objects, three hypotheses have been proposed: circumstellar (Keplerian) discs seen edge-on (e.g., \citealt{Norris1998, Minier2000}); outflows (e.g., \citealt{DeBuizer2003});  propagating shock fronts through rotating dense cores and ring-like structures (e.g., \citealt{Bartkiewicz2009}). The only way to discriminate among these different interpretations is deriving maser proper motions via multi-epoch VLBI observations.

The present paper focuses on the HMSFR \object{G16.59$-$0.05}.
In Sect.~2, we provide an up-to-date review of the single-dish and interferometric observations towards this region. Section~3 describes our VLBI observations of the 22.2~GHz H$_2$O, 6.7~GHz CH$_3$OH, and 1.665~GHz OH masers, together with the new Very Large Array (VLA) A-configuration observations of the radio continuum emission at 1.3 and 3.6~cm, which complemented VLA-C archival data at 0.7, 1.3, and 3.6~cm. Details of our data analysis are given in Sect.~4. In Sect.~5, we illustrate the spatial morphology, kinematics, and time-variability of individual maser species, and present results from our and archival VLA observations, constraining the properties of the radio continuum observed in correspondence of the masers.
Section~6 discusses the spatial association of the maser species and their overall kinematics, and draws a comprehensive picture of the phenomena observed in the HMSFR \G16 on angular scales from a few mas to tens of arcsec.  The main conclusions are summarized in Sect.~7.

\section{The HMSFR \G16}

The HMSFR \G16 (\object{IRAS 18182$-$1433}) has a near/far kinematic distance\footnote{We adopt a ``revised'' kinematic distance
using the prescription of \citet{Reid2009b}} of 4.4/11.9~kpc evaluated assuming a systemic velocity (V$_{sys}$) with respect to the local standard of rest (LSR) of \  59.9~\kms, inferred from observations of the NH$_3$~(3,3) and CH$_3$CN~(5--4) lines \citep{Codella1997,Furuya2008}. For consistency with previous
analyses of this source, in the following the near kinematic distance is adopted. At the near kinematic distance, the bolometric luminosity calculated from the IRAS fluxes is $2 \times 10^{4}$~L$_\odot$ (\citealt{Sridharan2002}) and the large-scale clump mass is  $\ge10^3$~M$_\odot$ (\citealt{Beuther2002a,Faundez2004,Hill2005,Furuya2008}).

On a large scale, mid-IR observations toward \G16 resolved the IRAS source in the region detecting 2 objects separated by about $10''$ along the NW--SE direction (\citealt{DeBuizer2005}; \citealt[][Fig.~1]{Furuya2008}). While the SE emission is associated with a MSX point source \citep{Walsh2003}, the fainter NW source ($\approx 0.2$~Jy at $18.1 \mu$m) is associated with weak, reddened NIR emission \citep{Testi1994} and was tentatively classified as a Class~\textrm{I} object \citep{DeBuizer2005}. The NW source is also coincident with the submillimeter (450 and 850~$\mu$m) continuum emission from the region, which may indicate that this is a younger, more embedded source than the one responsible for the SE emission (\citealt{Walsh2003,Williams2004,Williams2005}).
Single-dish surveys toward \G16 detected redshifted self-absorption in N$_2$H$^+$, HCO$^+$, H$^{13}$CO$^+$, and H$_2$CO rotational lines, which suggests the presence of inflow motions \citep{Fuller2005,Wu2007,Thomas2008}. Bolometer array observations at 1.2~mm \citep{Beuther2002a,Faundez2004} show a peak of dust emission in correspondence of the two mid-IR sources and an additional fainter source about $100''$ toward the SE, identified as a candidate High Mass Starless Core (HMSC) by \citet{Sridharan2005}.

High resolution observations at millimeter and radio wavelengths, both in line and continuum emission, revealed multiple sources within
the mid-IR region (see Fig.~\ref{fig8}, where we summarize both previous observations and our results).  A molecular core with a kinetic temperature of 54~K is identified through NH$_3$ observations by \citet{Codella1997}. Sensitive VLA-C observations at 3.6, 1.3~cm, and 7~mm detected two weak centimeter sources and one 7~mm continuum source, labeled ``c'', ``b'', and ``a'' by \citet{Zapata2006}, respectively. The centimeter source ``c'' appears to be associated with the strongest mid-IR source; the other two VLA sources
(with the 7~mm source ``a'' observed about $1''$ to the NW of the nearby centimeter source ``b'')
are offset by about $10''$ to the NW of the strongest mid-IR source, and correspond in position with the NW mid-IR emission.
Continuum measurements at 3~mm \citep{Furuya2008} and 1.3~mm \citep{Beuther2006} peak also at the same position as the VLA source ``a''. The spectral index suggests that the millimeter emission traces the main massive proto-stellar object in the field, whereas the centimeter feature ``b'' might indicate a thermal jet or, alternatively, an optically thin H~\textsc{ii} region \citep{Zapata2006}.
Emission in several molecular lines typical of HMCs was detected at an intermediate position between source ``a'' and ``b'' \citep{Beuther2006,Furuya2008}. The rotational temperature inferred from the CH$_3$CN lines is about 130--150~K in this region, consistent with the presence of an HMC.

In this work we will focus our attention on the NW sources, namely the HMC and the associated sources ``a'' and ``b'', hosting the maser emissions detected toward \G16. OH masers were reported in the main 1.665~and~1.667~GHz, and the satellite 1.720~GHz lines \citep{Caswell1983a,Edris2007}, whereas the 1.612~GHz OH satellite was observed in absorption. CH$_3$OH thermal lines were observed at 25, 96.7, 157, 241.7, and 255~GHz by \citet{Leurini2007}, and maser emission was reported for the  Class~\textrm{II} 6.7~GHz \citep{Menten1991,Caswell1995a,Szymczak2000} and 12.2~GHz \citep{Caswell1995b}, and the Class~\textrm{I} 44.1~GHz \citep{Slysh1994} and 95.2~GHz \citep{Valtts2000} transitions.
High-resolution ($\geq1''$), interferometric observations towards \G16 of the 22.2~GHz H$_2$O \citep{F&C1999,Beuther2002c}, 1.665~GHz OH \citep{F&C1999}, and 6.7~GHz CH$_3$OH \citep{Walsh1998} maser lines, associate the maser activity with the sources ``a'' and ``b''.

A wide angle, massive ($\geq30$~M$_\odot$) bipolar outflow emerging from the HMC was detected through $^{12}$CO and CO isotopologues rotational lines (1--0 and 2--1) at an angular resolution of about $10''$ by several authors \citep{Beuther2002b,Furuya2008,LopezSepulcre2009}. At the higher spatial resolution ($\approx 3''$) of the SMA interferometer, the $^{12}$CO~(2--1) emission was resolved in a multiple outflow system \citep{Beuther2006}: the strongest and better collimated lobes are oriented along two, approximately perpendicular directions, NW--SE and NE--SW. The outflow system is centered on the position of the millimeter source. Furthermore, VLA SiO~(1--0) observations revealed emission elongated toward the N--S direction, which could be suggestive of an additional third outflow \citep{Beuther2006}.

The large collection of multi-wavelength data reported here suggests that the \G16 region is an active site of multiple, massive star formation.

\section{Observations and Calibration}

\subsection{Archival Data and VLA continuum}

We observed G16.59--0.05 with the VLA\footnote{The VLA is operated by the National Radio Astronomy Observatory (NRAO). The NRAO is a facility of the National Science Foundation operated under cooperative agreement by Associated Universities, Inc.} in the C- and A-array configurations
at X band, and in the A-array configuration at K band. The VLA--C and VLA--A observations
were made respectively in April and October 2008. The source had already
been observed at K~and~Q~bands with the VLA--C by Zapata et al. (2006). We
retrieved their data from the VLA archive (Program Code: AZ149)
and re-reduced them to improve on the angular resolution of the maps presented
in their paper. In particular, we imaged the Q-band data using natural weighting,
achieving an angular resolution about a factor~2 better than that of the tapered
maps by Zapata et al. (2006).
At 3.6~cm we used the continuum mode of the correlator, resulting in an
effective bandwidth of 172~MHz. For the 1.3~cm observations, we used mode
``4'' of the correlator, with a pair of 3.125~MHz bandwidths (64 channels) centered on the strongest H$_2$O
maser line and a pair of 25~MHz bandwidths (8 channels) sufficiently offset
from the maser lines to obtain a measurement of the continuum emission.
This mode was used to improve on the relative position accuracy between
the maser emission and the high-resolution observations of
the 1.3~cm continuum.
The two bandwidths centered at the same frequency (measuring the two circular polarizations)
were averaged.

At X-band, 3C\,286 (5.2~Jy) and 3C\,48 (3.1~Jy) were used as primary flux
calibrators, while 1832--105 (1.4~Jy) was the phase calibrator.  For the
K-band observations, the primary flux calibrator was 3C\,286 (2.5~Jy), the
phase calibrator 1832--105 (1.0~Jy), and the bandpass calibrator
1733-130 (3.7~Jy).

The data were calibrated with the NRAO AIPS software package using standard procedures. Only for the
VLA--A data at 1.3~cm, several cycles of self-calibration were applied to
the strongest maser channel, and the resulting phase and amplitude
corrections were eventually transferred to all the other line channels and to
the K-band continuum data. This procedure resulted in a significant (at least
a factor 2) improvement of the signal-to-noise ratio (SNR).

\subsection{Maser VLBI}

We conducted VLBI observations of the H$_2$O and CH$_3$OH masers (at several epochs), and of the OH maser (single epoch) toward \G16 in the K, C, and L bands, respectively. To determine the maser absolute positions, we performed phase-referencing observations by fast switching between the maser source and the calibrator J1825$-$1718. This calibrator has an angular offset from the maser source of $2\fdg9$ and belongs to the list of sources defining the International Celestial Reference Frame (ICRF). Its absolute position is known to better than $\pm2$~mas and its flux measured with the Very Long Baseline Array (VLBA) at S and X bands is 48 and 112~m\Jyb, respectively \citep{Fomalont2003}.
Five fringe finders (J1642+3948; J1751+0939; J1800+3848; J2101+0341; J2253+1608) were observed for bandpass, single-band delay, and instrumental phase-offset calibration.

Data were reduced in AIPS employing the VLBI spectral line procedures.
Visibility amplitudes were calibrated applying the information on the system temperature and antenna sensitivity provided by the monitoring procedure at each VLBI antenna site.
A priori delay and phase calibration for feed rotation, correction for ionospheric delays (AIPS task TECOR), and for inaccuracies of the extrapolated Earth's orientation parameters used at the time of correlation (AIPS task CLCOR), were applied to the data set (e.g., EVN data analysis guide\footnote{$www.evlbi.org/user_-guide/guide/user_-guide.html$}).
A posteriori phase calibration for short-term atmospheric fluctuations and phase-reference structure was derived by fringe-fitting and self-calibrating the strong, compact, emission in a given maser channel. Applying this calibration, visibilities of all maser channels were referred in phase (i.e. in position) to the emission centroid of the reference maser channel. The
determination of the absolute position of the reference maser channel was accomplished in either of the two following ways: imaging the calibrator J1825$-$1718 data after applying corrections derived working with the reference maser channel (\emph{inverse phase-referencing}); or, by performing the reverse procedure, imaging the reference maser channel by applying the fringe-fitting and self-calibration solutions from the calibrator data (\emph{direct phase-referencing}). In the inverse phase-referencing procedure, before fringe-fitting the maser data, the visibility phase center was shifted (AIPS tasks CLCOR and UVFIX) close to the effective maser position, in order to avoid degradation of the calibrator image (e.g., \citealt{Reid2009}). For well detected (maser and calibrator) signals (SNR $\ge$ 10), the two procedures always gave consistent results.

\subsubsection{VLBA observations: 22.2~GHz H$_2$O masers}

\begin{figure*}
\centering
\includegraphics[width=7cm]{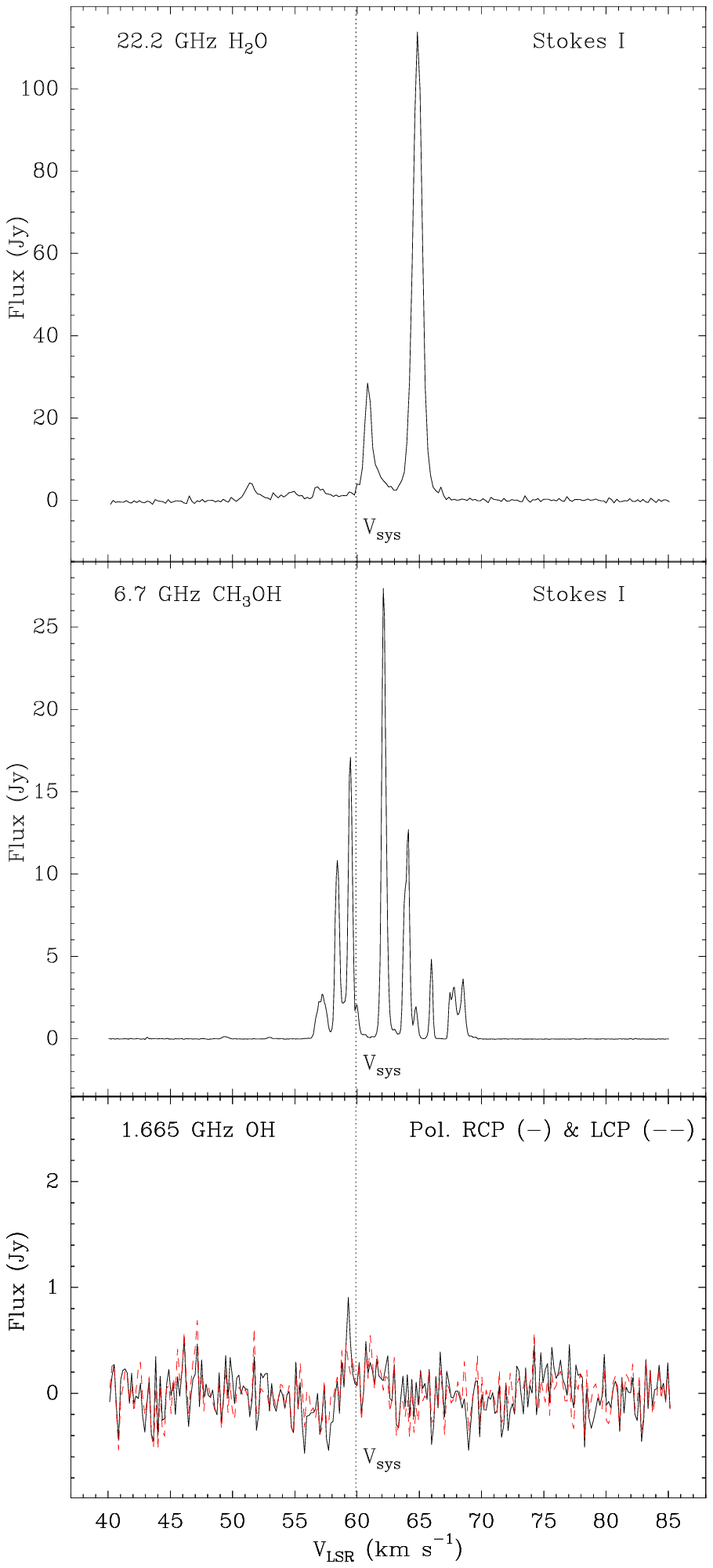}
\caption{Total-power spectra of the H$_2$O, CH$_3$OH and OH masers toward \G16. \emph{Upper panel:} system-temperature (T$_{sys}$) weighted average of the 22.2~GHz total-power spectra of the 9 VLBA antennas observing on 2007 January 4 (BR, HN, KP, LA, MK, NL, OV, PT, SC). \emph{Middle panel:} Effelsberg total-power spectrum of the 6.7~GHz methanol maser emission on 2007 March 16. \emph{Lower panel:} T$_{sys}$-weighted average of the 1.665~GHz total-power spectra of the 10 VLBA antennas observing on 2007 April 13. Continuous (black) and dashed (red) lines are used to denote the right (RCP) and left (LCP) circular polarizations, respectively. The vertical dotted line across the spectra indicates the systemic velocity (V$_{sys}$) inferred from CH$_3$CN measurements.}
\label{fig1}
\end{figure*}

We observed the HMSFR \G16 (tracking center: R.A.(J2000)~$=18^h21^m09\fs11$ and Dec.(J2000)~$=-14\degr31'48\farcs5$) with the VLBA\footnote{The VLBA is operated by the NRAO.} in the $6_{16}-5_{23}$ H$_2$O transition (rest frequency 22.235079~GHz). The observations (program code: BM244) consisted of 4 epochs: on April 9, June 28, and September 18, 2006, and on January 4, 2007.
During a run of about 6~h per epoch, we recorded the dual circular polarization through a 16~MHz bandwidth centered on a LSR velocity of 60.0~\kms. The data were processed with the VLBA FX correlator in Socorro (New Mexico) using an averaging time of 1~s and 1024 spectral channels. The total-power spectrum of the 22.2~GHz masers toward \G16 is shown in Fig.~\ref{fig1} (top panel). This profile was obtained by averaging the
total-power spectra of all VLBA antennas, weighting each spectrum with the antenna system temperature (T$_{sys}$).

The natural CLEAN beam was an elliptical Gaussian with a FWHM size of about $1.4~\textrm{mas} \times 0.4~\textrm{mas}$ at a P.A. of $-16\degr$ (east of north), with little variations from epoch to epoch. The interferometer instantaneous field of view was limited to about $2\farcs7$ .
At the various observing epochs, using an on-source integration time of about 2.5~h, the effective rms noise level of the channel maps ($\sigma$) varied in the range \ 0.01--0.04~\Jyb. The spectral resolution was 0.2~\kms.

\subsubsection{VLBA observations: 1.665~GHz OH masers}

We observed the HMSFR \G16 (tracking center: R.A.(J2000)~$=18^h21^m09\fs21$ and Dec.(J2000)~$-14\degr31'48\farcs3$)  with the VLBA in the $^2\Pi_{3/2}$ $\rm{J}=3/2$ OH transition (rest frequency 1.665401~GHz), on  April 13, 2007 (program code: BM244M). During a run of about 6~h, we recorded the dual circular polarization through two bandwidths of 1~MHz and 4~MHz, both centered on a LSR velocity of 60.0~km~s$^{-1}$.
The 4~MHz bandwidth was used to increase the SNR of the weak L-band signal of the continuum calibrator.
The data were processed with the VLBA FX correlator in two correlation passes using either 1024 or 512 spectral channels for the 1~MHz and 4~MHz bands, respectively. In each correlator pass, the data averaging time was 2~s. The T$_{sys}$-weighted mean of antenna total-power spectra for the right and left circular polarizations at 1.665~GHz are shown in Fig.~\ref{fig1} (bottom panel).

The natural CLEAN beam was an elliptical Gaussian with a FWHM size of $24~\textrm{mas} \times 10~\textrm{mas}$ at a P.A. of $3\degr$. The interferometer instantaneous field of view was limited to about $18\farcs5$.
With an on-source integration time of about 1.9~h, the effective rms noise level on the channel maps was about 0.01~\Jyb. The 1~MHz band spectral resolution was 0.2~\kms.

\subsubsection{EVN observations: 6.7~GHz CH$_3$OH masers}

We observed the HMSFR \G16 (tracking center: R.A.(J2000)~$=18^h21^m09\fs13$ and Dec.(J2000)~$=-14\degr31'48\farcs5$) with the European VLBI Network (EVN)\footnote{The European VLBI Network is a joint facility of European, Chinese and other radio astronomy institutes founded by their national research councils.} in the $5_{1}-6_{0} A^+$ CH$_3$OH transition (rest frequency 6.668519~GHz).
This work is based on 3 epochs (program codes: EM061, EM069), separated by about 1~yr, observed on February 26, 2006, on March 16, 2007, and on March 15, 2008.
At the first two epochs, antennas involved in the observations were Cambridge, Jodrell2, Effelsberg, Hartebeesthoek, Medicina, Noto, Torun and Westerbork. Since the longest baselines involving the Hartebeesthoek antenna (e.g., the Ef-Hh baseline is about 8042~km) heavily resolve the maser emission and do not produce fringe-fit solutions, the Hartebeesthoek antenna was replaced with the Onsala antenna in the third epoch. During a run of about 6~h per epoch, we recorded the dual circular polarization through two bandwidths of 2~MHz and 16~MHz, both centered on a LSR velocity of 60.0~\kms.
The 16~MHz bandwidth was used to increase the SNR of the weak continuum calibrator.
The data were processed with the MKIV correlator at the Joint Institute for VLBI in Europe (JIVE - Dwingeloo, The Netherlands) using an averaging time of 1~s and 1024 spectral channels for each observing bandwidth.
The Effelsberg total-power spectrum at 6.7~GHz toward \G16 is shown in Fig.~\ref{fig1} (middle panel).

The natural CLEAN beam was an elliptical Gaussian with a FWHM size of about $15~\textrm{mas} \times 5~\textrm{mas}$ at a P.A. of $30\degr$, with little variations  from epoch to epoch. The interferometer instantaneous field of view was limited to about $9\farcs2$.
At each observing epoch, using an on-source integration time of about 2.2~h, the effective rms noise level of the channel maps varied in the range \ 0.008--0.1~\Jyb. The 2~MHz band spectral resolution was 0.09~\kms.

\section{Data Analysis}\label{spot_feature}

Mapped maser channels were searched for emission above a conservative threshold, equal to the absolute value of the minimum in the map, typically corresponding to values greater than 5--7$\sigma$. Parameters (position, intensity, flux, and size) of the detected maser (and calibrator) emission have been derived by fitting a two-dimensional elliptical Gaussian. The term ``\emph{spot}'' is used to refer to maser emission on a single channel map. We checked the contour plot of each detected spot and repeated the Gaussian fit to optimize the number of  Gaussian components in case of complex spatial structure. We took particular care to
identify and remove spurious spots owing to side-lobe effects of the synthesized beam.
The uncertainty of the relative fit position is estimated using the expression:
$ (\Delta \rm{\theta})^2 = (0.5 \, {\rm FWHM} / {\rm SNR})^2 + (50  \rm {\mu as})^2 $.
The first contribution represents the Gaussian fit uncertainty \citep{Reid1988}, with FWHM being the (undeconvolved) full-width half-maximum size of the spot, and SNR the signal-to-noise ratio of the Gaussian fit. The second contribution represents a noise level added to take into account position uncertainties associated with correlator-unmodeled signal propagation through the troposphere \citep{Pradel2006}.

The VLBI angular and velocity resolutions are sufficiently high to resolve the (spatial and velocity) structure of single masing-clouds.
We collected spots observed on contiguous channel-maps and spatially overlapping (within their FWHM size), into a single maser ``\emph{feature}''.
The position and LSR velocity of a given feature are estimated from the error-weighted mean position and the intensity-weighted mean LSR velocity of its spots, respectively.
If, for a given feature, spot position errors are comparable with the size of the spot distribution (as it is the case of most 22.2~GHz H$_2$O masers, extended for a few tenth of mas), the feature's position uncertainty is evaluated by the error-weighted standard deviation of the spots positions.
For extended maser features (as most 6.7~GHz CH$_3$OH masers, extended for a few mas), for which the diameter of the spot distribution is significantly larger than the spot position uncertainties, the position error is taken equal to the error-weighted mean of the spot position errors.
The uncertainty in the absolute positions of the maser features is calculated by taking the square mean of three independent error terms:
1) the feature's relative-position uncertainty; 2) the position error of the (calibrator or reference maser) signal in the phase-referenced maps; 3) the position uncertainty of the calibrator J1825$-$1718. This latter term dominates the absolute-position error for water and methanol maser features.

We have used two main criteria to establish the correspondence of maser features over the epochs: 1) persistence of the relative distribution of a group of features (see inset in Fig.~\ref{fig3};  see also \citealt{Goddi2006}); 2) assumption of uniform motions (e.g., \citealt{Reid1988,Gwinn1992}). The tolerance in the change of peak LSR velocity for persistent features was fixed to less than their spectral FWHM (see Fig.~\ref{fig6}). Relative proper motions for features persisting over (at least) 3 epochs, have been calculated by performing a  linear fit of their varying (relative) position with time. The derived proper motions are a measure of the feature mean motion over a time baseline of (up to) 1~yr, for water, and 2~yr, for methanol masers.

Absolute velocities are derived by adding to the relative velocities the absolute motion of the reference maser feature (the one to which velocities
have been referred), determined with respect to the calibrator J1825$-$1718. To estimate maser motions relative to the LSR reference frame of the HMSFR under study,  the measured absolute motion of the reference maser feature has to be corrected for the apparent proper motion due to the earth revolution around the Sun (parallax), the Solar Motion and the differential Galactic Rotation between our LSR and that of the
HMSFR. Taking the IAU values for the Solar Motion and the Galactic Rotation (R$_0$ = 8.5~kpc, $\Theta_0$ = 220~\kms), and assuming the near kinematic distance of 4.4~kpc,  we derive an apparent motion of 12~\kms to the west and 38~\kms to the south. These values can be affected by
large errors of 10--20~\kms, owing to uncertainties in the Solar and Galactic standard values, as well as to the recent observational evidence that
HMSFRs might be orbiting the Galaxy about 15~\kms slower than expected for circular orbits \citep{Reid2009b}. Absolute velocities have been
estimated for water masers only, whose measured relative motions (see Sect.~\ref{h2o_results}) are found on average significantly larger than the expected uncertainty of the correction for the apparent motion.

\section{Results}

\subsection{Radio continuum emission}

In the following, we focus our attention on the centimeter source labeled ``b'' by \citet{Zapata2006}, the one located nearby the millimeter continuum, since it appears to be spatially associated with the observed 6.7~GHz CH$_3$OH and 22.2~GHz H$_2$O maser distributions in \G16 (Fig.~\ref{fig3}). Our VLA--A and VLA--C measurements of the 1.3~and~3.6~cm continuum, and the VLA--C archival data at 0.7~and 1.3~cm, are summarized in Table~\ref{tab1}.
Contour plots of the 1.3~cm and 7~mm continuum emission are presented in Figs.~\ref{fig3}--\ref{fig5} and~\ref{fig7}.
At 1.3~and~3.6~cm, the emission is resolved out by the VLA--A and we were not able to accurately calibrate the relative position of the continuum with respect to the H$_2$O masers. On the other hand, the radio continuum appears as a compact (or marginally resolved) source with the VLA--C and the peak positions of the 1.3~and~3.6~cm emissions coincide with one another within the uncertainties. In agreement with \citet{Zapata2006}, the peak of the 7~mm continuum is found offset by $1''$--$2''$ to the NW with respect to the 1.3~and 3.6~cm emission peak. The 7~mm maps of \citet[][Figs.~4~and~5]{Zapata2006}, with a synthesized beam of about
1\farcs7, show an unresolved spur of emission pointing to the SE toward the 1.3~cm continuum.
We did not use their tapering of the data and thus attained  at 7~mm a synthesized beam of about 0\farcs7. With this resolution, the emission spur is resolved into a double lobe structure elongated in the SE--NW direction. This extended emission is detected at a low-significance level of only 3-4$\sigma$, but the fact that is also observed in the tapered maps of \citet{Zapata2006} indicates that it is real.
Figures~\ref{fig3}--\ref{fig5} show that the 7~mm southeastern lobe is found at a position close to the peak of the 1.3~cm (and~3.6~cm) continuum.
We consider the small offset (between 0\farcs1--0\farcs2) of the 7~mm southeastern lobe  with respect to the 1.3~cm continuum peak not to be real, but being an artifact owing to the low SNR of this 7~mm emission feature. In the following discussion, we assume a single source  to be
responsible for the emissions at 1.3~cm, 3.6~cm, and from the 7~mm southeastern lobe, and denote this source with the label ``b1'' (see Table~\ref{tab1}).
Figure~\ref{fig2} shows the spectral energy distribution (SED) of the radio emission of  source ``b1''.
In Table~\ref{tab1}, the 7~mm lobe, offsets to the NW of ``b1'' by about 0\farcs5--0\farcs7, is indicated with the label ``b2''.

\subsection{22.2~GHz H$_2$O masers}\label{h2o_results}

We searched for water maser emission through the whole field of view, $1'' \times 1''$, and range of LSR velocities, from 50 to 68~\kms, explored by \citet{Beuther2002c} using the VLA--B configuration. The centroid of their water maser distribution falls inside the area of our detections.

We have detected 40 distinct water maser features, distributed within an area of about $0\farcs4 \times 0\farcs8$ (Fig.~\ref{fig3}b). Most of the maser emission comes from 35 features distributed across a region of size $\sim350$~mas to the west of the 1.3~cm continuum peak.
A cluster of 5 features spread over $\sim30$~mas is observed offset by about 0\farcs6 to the north of the 1.3~cm continuum source.
The individual features properties are presented in Table~\ref{wat_tab}. Maser intensities range from 0.10 to 65.49~Jy~beam$^{-1}$.
14 features (35\% of the total) persisted over at least 3 epochs,  9 of which lasted 4 epochs.
The spread in LSR velocities ranges from 67.9~\kms, for the most redshifted feature (\#~23), to 51.4~\kms, for the most blueshifted one (\#~9). Water maser absolute positions and LSR velocities are plotted in Fig.~\ref{fig3}b. The inset of Fig.~\ref{fig3}b shows the details
of a persistent structure of water maser emission, which presents a regular variation of LSR velocities with position, from 62~to~67~\kms across a region of about 45~mas. This structure harbors several among the strongest maser features, including feature~\#~1 which experienced, between the first and second epoch, a powerful maser flare.
Using a time baseline of 9~months, we have measured relative transverse velocities of water maser features with a mean accuracy of about 30\%. The amplitude of relative proper motions ranges from 15.1$\pm$4.1~\kms, for feature~\#~22, to 117.1$\pm$7.0~\kms, for feature~\#~25.
Figure~\ref{fig5}a presents relative positions and velocities of the features, determined with respect to the bright, slowly-variable ($\lesssim 10\%$) feature~\#~4. Both the direction and the uncertainty of the relative proper motions are indicated.
Figure~\ref{fig5}b shows the derived absolute proper motions of the persistent water maser features.
Since absolute proper motions are affected by large uncertainties, we show only the inferred mean direction of the motion.

\subsection{1.665~GHz OH masers}\label{oh_results}

\citet{F&C1999} observed the HMSFR \G16 with the VLA  (HPBW of $7\farcs2 \times 1\farcs0$) in the LCP band and reported 7 distinct centers of 1.665~GHz OH maser emission. Their brightest detection had a flux density of 2.2~Jy at the V$_{\rm LSR}$ of 58.9~\kms.
Figure~\ref{fig4} shows the area of the 1.665~GHz OH maser detections by \citet{F&C1999}.

No 1.665~GHz OH maser signal is detected on single VLBA baselines over 2~min scans, with an estimated rms sensitivity of 0.9~Jy. The mean
total-power spectrum of the 10 VLBA antennas (Fig.~\ref{fig1}) shows a tentative detection in the RCP band of a 0.85~Jy line at a LSR velocity of about 59~\kms. By imaging maser data after phase-referencing to the calibrator J1825$-$1718, no signal was detected above a 5$\sigma$ level of 0.06~\Jyb, across an area of about $ 12'' \times 12'' $ centered on the VLA 1.665~GHz OH maser position reported by
\citet{F&C1999}. Since it is unlikely that the VLA OH maser position is wrong by more than 6\arcsec, we consider residual
ionospheric phase errors as the most probable cause of image degradation and non-detection of the maser signal.

\subsection{6.7~GHz CH$_3$OH masers}\label{ch3oh_results}

We covered the range of LSR velocities (from 52 to 69~\kms) over which 6.7~GHz maser emission was detected in the high sensitivity, Parkes-64~m observations by \citet{Caswell1995a}, and imaged the whole field of view ($0\farcs23 \times 0\farcs27$) where CH$_3$OH  masers were detected by \citet{Walsh1998} using the ATCA interferometer. The centroid of the ATCA 6.7~GHz maser distribution falls inside the area  of our maser detections.

We have detected 39 distinct 6.7~GHz CH$_3$OH maser features, most (95\%) of which distributed over an area of about $0\farcs4 \times 0\farcs4 $  centered on the 1.3~cm continuum peak. Hereafter we refer to this group of 6.7~GHz maser features as the ``main cluster''. Figure~\ref{fig3}a shows the spatial distribution and the LSR velocities of the 6.7~GHz masers.
The individual feature properties are presented in Table~\ref{met_tab}.
Two features, labeled \#~28~and~38, are found offset from the
main cluster to the west and the north, respectively, and fall close (within about 30~mas) to water maser features with also a good correspondence in LSR velocity (within 1.5~\kms; see Fig.~\ref{fig4}). The spread in LSR velocities ranges from 68.5~\kms, for the most redshifted feature (\#~8), to 52.9~\kms, for the most blueshifted one (\#~38). The main cluster presents a bipolar distribution with a clear separation of l.o.s velocities: redshifted features toward the NW, and blueshifted ones toward the SE. The average l.o.s. velocities over the NW
and SE regions are  \ $+3.7$ and \ $-1.5$~\kms, respectively.  Maser intensities range from 0.06 to 20.62~\Jyb.
25 features (64\% of the total) persisted over the 2~yr interval covered by our 3 observing epochs. For each maser feature, Table~\ref{met_tab} reports the mean brightness variability defined as the ratio between the variation of the brightness (of the strongest spot) and the average of the minimum and maximum brightness. Relative positions are given with respect to the structurally-stable, compact, and bright feature~\#~2, whose intensity varied less than about 8\% (see Fig.~\ref{fig6}). The mean brightness variability of methanol maser features was less than about 22\%, spanning a range from 2\%, for feature~\#~16, to about 45\%, for feature~\#~21.  These estimates of brightness variability should be taken as upper limits since they include effects from the slightly varying beam shape among different epochs as well as amplitude calibration uncertainties.

We have calculated the geometric center (hereafter ``center of motion'', identified with label~\#~0 in Table~\ref{met_tab}) of features with a stable spatial and spectral structure persisting over the 3 observing epochs. We refer our measurement of transverse motions to this point (Fig.~\ref{fig7}). Using a 2~yr time baseline, relative transverse velocities of individual maser features are determined with a mean accuracy of better than 30\%. Proper motions, derived only for maser features with a stable (spatial and spectral) structure, are calculated by performing a linear fit of feature position vs. time. An analysis of deviations from linear motion, to establish the existence of possible accelerated motions, is postponed after an upcoming fourth epoch.  The absolute distribution and relative proper motions of methanol 6.7~GHz masers are plotted in Fig.~\ref{fig7}.
The amplitude of relative proper motions ranges from $3.6\pm1.4$~\kms, for feature~\#~5, to $11.2\pm1.6$~\kms, for feature~\#~6. The mean transverse velocity is equal to 6.2~\kms, about 2--5 times greater than the spread in l.o.s. velocities.
It is interesting to note that most of the transverse velocities for features placed in the NW of the main cluster are oriented towards N or NE, whereas features in the SE of the main cluster move towards S or SW.

\subsubsection{Interpretation of 6.7~GHz maser proper motions}\label{meth_kinematics}

In order to interpret correctly the proper motion of a maser feature we need to establish if the measured motions are real or apparent. Change in physical and/or excitation conditions of maser-emitting gas can in principle lead to apparent motions (Christmas-tree effect). Concerning 22.2~GHz H$_2$O masers, accurate proper motion measurements towards several HMSFRs strongly suggest that this maser emission traces physical motions of gas bullets, powered
by stellar winds or jets emitted from massive YSO(s) (e.g., \citealt{Goddi2006}).  Since our study is one of the first attempts to use proper motions of 6.7~GHz methanol masers to infer gas kinematics in HMSFRs, we propose a few arguments supporting the kinematic interpretation of features' proper motions.

Figure~\ref{fig6} shows the time evolution during 2 years (from February 2006 to March 2008) of the morphology and spectral structure of three persistent 6.7~GHz maser features (\# 2, 8, and 16), selected to be representative of the observed feature properties.
For each feature and each observing epoch, three sets of data are presented: the spectral profile (\emph{left panel of Fig.~\ref{fig6}});
the image of the most extended feature emission obtained mapping the weakest channels in the spectral wings (hereafter the feature ``image''; \emph{middle panel of Fig.~\ref{fig6}}); the spatial and LSR velocity distribution of the brightest spots corresponding to the feature (hereafter the ``internal velocity gradient''; \emph{right panel of Fig.~\ref{fig6}}).
To compare images of the same feature among different epochs, we cleaned the visibility data using the same restoring beam.

The slightly resolved shape of the feature image remains approximately constant across the epochs. That is essentially a consequence of the remarkable persistency of the feature spectrum and spot relative positions, with internal velocity gradients of typically $\sim0.1$~\kms~mas$^{-1}$ (see Fig.~\ref{fig6}). This might be interpreted as the maser emission at different epochs coming from the same blob of gas, whose internal (physical and kinematical) properties  do not change significantly over time.
On the contrary, if proper motions were due to different regions of the cloud being excited at consecutive times, one would not expect
the internal velocity and geometrical structure of the feature to be so well preserved. Forthcoming VLBI epochs will allow to establish the persistency of the feature spectrum and internal velocity gradient over a longer time baseline.

Another indication that we are measuring real motions comes from
the smooth variation of the observed 6.7~GHz maser proper motions with feature position. Looking at Fig.~\ref{fig7} one notes that proper
motions of nearby features have similar orientations and amplitudes, which suggests that masers trace a smoothly varying  velocity field. In
contrast, the Christmas-tree effect is expected to produce apparent motions uncorrelated with positions. Finally, considering \emph{the whole} distribution of measured 6.7~GHz proper motions,  it is quite evident that the velocities in the plane of the sky of redshifted and blueshifted features are antiparallel and perpendicular to the line connecting them (see Fig~\ref{fig7}). In the next section, we interpret the whole distribution of l.o.s. and transverse velocities in terms of rotation around the center of motion of the 6.7~GHz masers.

\section{Discussion}

The simplest interpretation for the bipolar distribution of 6.7~GHz maser l.o.s. velocities and the regular pattern of observed transverse velocities is that the masers are tracing rotation around the center of motion (see Fig.~\ref{fig7}). Since transverse velocities are generally larger than l.o.s. velocities, most of the features move close to the plane of the sky.
The bipolar distribution of l.o.s. velocities indicates that the rotation axis is inclined with respect to the line of sight, with redshifted (blueshifted) features to the NW (SE) of the main cluster rotating away from (towards) the observer. Taking the average inclination with the plane of the sky of all the measured proper motions, the rotation axis should form an angle of about 30\degr\ with
the l.o.s. . Assuming centrifugal equilibrium the required central mass is 35~M$_{\odot}$. This dynamical mass ($\rm M_{dyn}=R V_{rot}^2/G$) has been obtained taking the mean velocity amplitude, $\rm V_{rot} = 7.2$~\kms, and the average distance from the center of motion, $\rm R = 600$~AU, of the 6.7~GHz maser features with measured proper motions.  Although the derived value is affected by  large errors owing to the uncertain source distance and
 the poor knowledge of the maser geometry, it is consistent with the 6.7~GHz masers being commonly found associated with massive YSOs.
The position angle ($\approx-45\degr$) of the main cluster of 6.7~GHz masers is approximately perpendicular to the NE--SW direction of the less collimated $^{12}$CO~(2--1) outflow detected by \citet{Beuther2006}. The NE--SW elongated spur visible in the NH$_3$~(2,2) map of Fig.~\ref{fig4} might also mark dense gas close to the axis of this outflow.  The elongated distribution of 6.7~GHz masers could then trace gas in a
disk/toroid rotating around the massive YSO which drives the NE--SW $^{12}$CO~(2--1) outflow (sketch in Fig.~\ref{fig8}).

\begin{figure*}
\centering
\includegraphics [angle=-90,width=7cm]{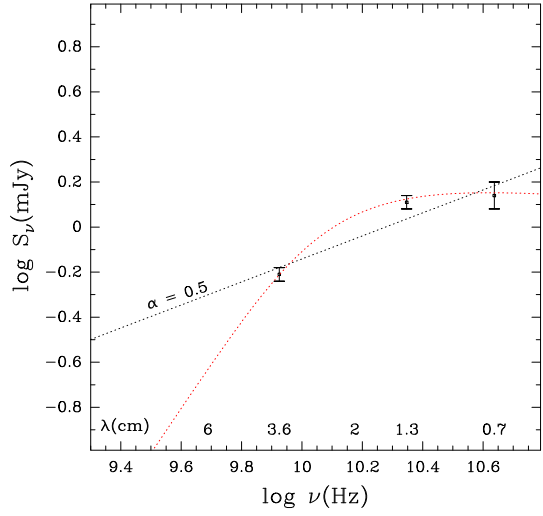}
\caption{Spectral energy distribution of the VLA component ``b1'' toward the HMSFR  \G16.
Dots and errorbars report the values and the associated errors (1$\sigma$) of the fluxes at 0.7, 1.3 and 3.6~cm measured with the VLA--C  (see Table~\ref{tab1}). The derived spectral index, indicated by the black dotted line, is \ $\alpha$ = 0.5. The red dotted line shows the best fit of the measured fluxes with a model of an homogeneous H~\textsc{ii} region ionized by a Lyman continuum of \ 3 $\times$ 10$^{45}$~s$^{-1}$ with a radius of about 0\farcs04.}
\label{fig2}
\end{figure*}

\begin{figure*}[htbp]
\centering
\includegraphics[angle=0.0,scale=1.0]{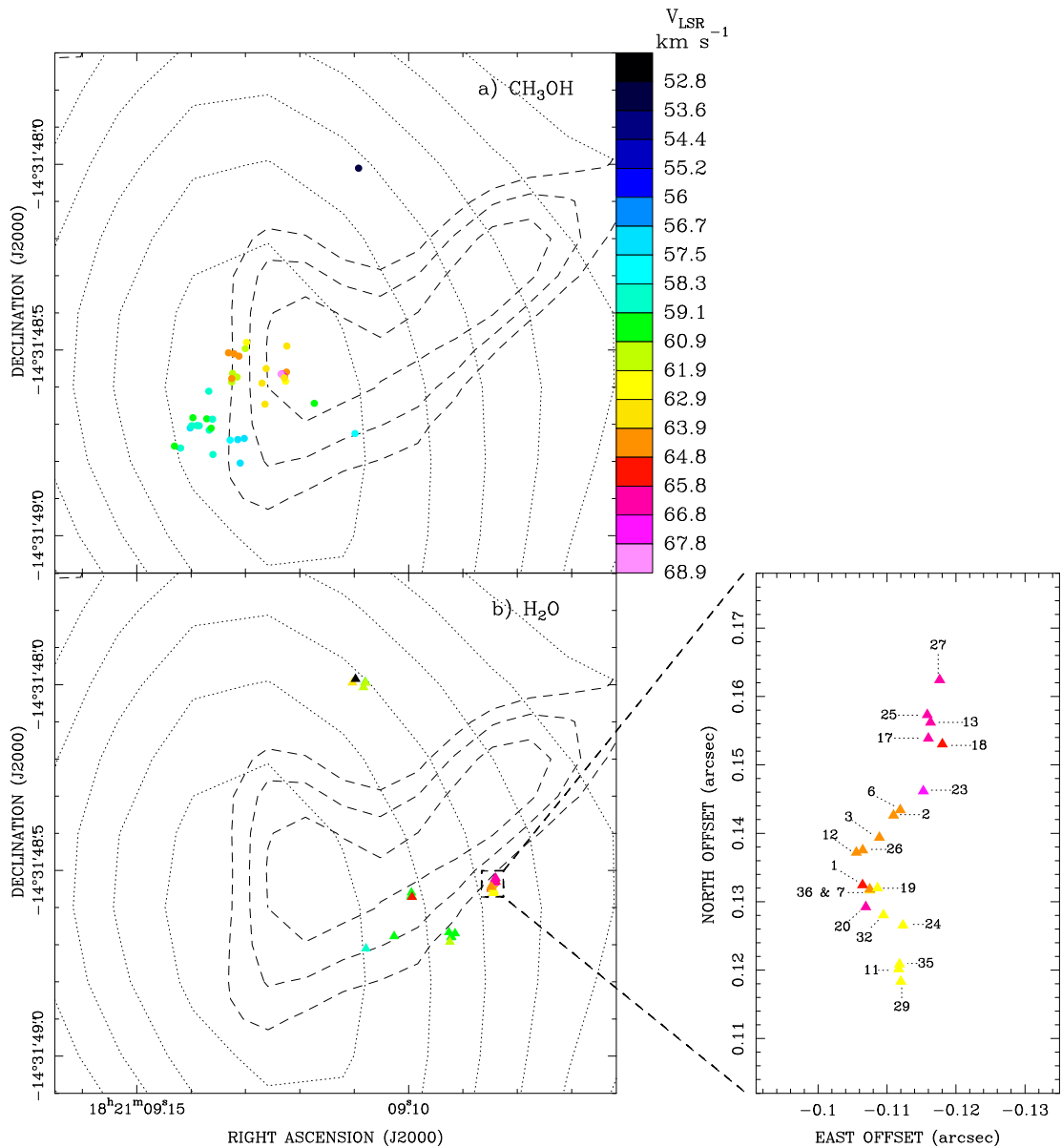}
\caption{Absolute positions and LSR velocities of maser species observed in \G16: \emph{a)} CH$_3$OH (dots); \emph{b)} H$_2$O (triangles). Different colors are used to indicate the maser LSR velocities, according to the color scale on the right-hand side of the plot, with green representing the systemic velocity of the HMC.  The VLA 1.3~cm and 7~mm continuum emissions are plotted with dotted and dashed contours, respectively. The 1.3~cm contour levels range from 40 to $90\%$ of peak emission (0.6~m\Jyb) at multiples of $10\%$, and from 70 to $90\%$ for the 7~mm peak (0.7~m\Jyb). \emph{Inset:} enlargement of the N--S threadlike structure described in Sect.~\ref{h2o_results}. Maser features are indicated with the label numbers given in Table~\ref{wat_tab}.}
\label{fig3}
\end{figure*}

\begin{figure*}[htbp]
\centering
\includegraphics[angle=0.0,scale=1.3]{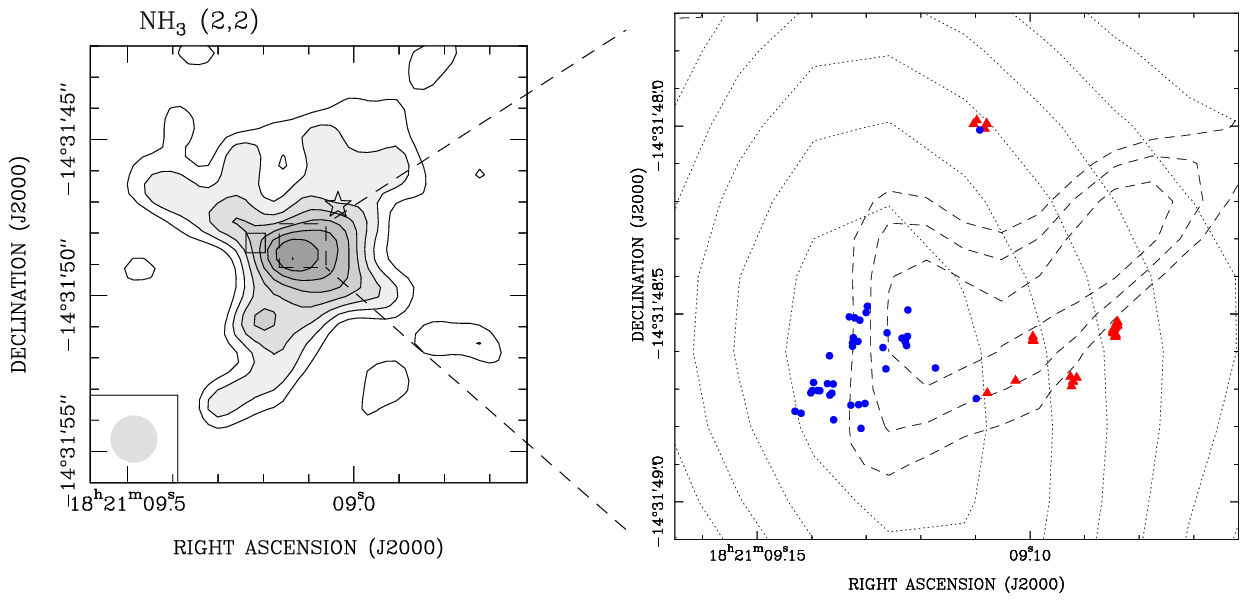}

\caption{Collection of the sub-arcsec observations toward \G16. \emph{Left panel:} map in the NH$_3$ (2,2) line from \citet{Codella1997}. Contour levels range from 30 to $90\%$ of the peak emission (38~m\Jyb) at multiples of $10\%$. The restoring beam is shown in the lower left corner of the panel. The star marks the peak position of the 7~mm continuum emission imaged by \citet{Zapata2006}. The empty square indicates the area where OH masers were detected by \citet{F&C1999}.
\emph{Right panel:} enlargement of the region over which we have detected maser emission. Red triangles and  blue dots represent H$_2$O and CH$_3$OH maser positions from our VLBI measurements, respectively. The VLA 1.3~cm and 7~mm continuum emissions are given with dotted and dashed contours, respectively, using the same contour levels shown in Fig.~\ref{fig3}.}
\label{fig4}
\end{figure*}

\begin{figure*}[htbp]
\centering
\includegraphics[angle=0.0,scale=1.0]{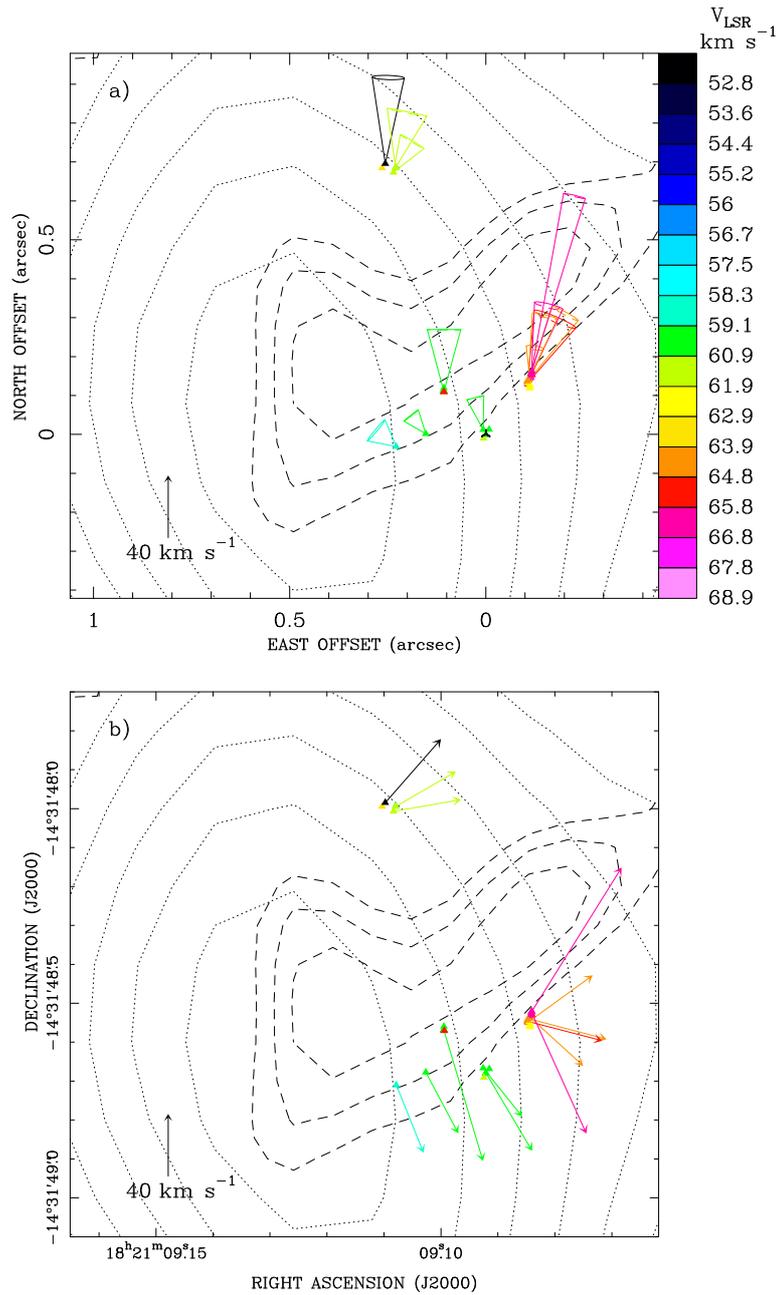}
\caption{22.2~GHz H$_2$O maser  kinematics toward \G16. \emph{a)} positions (triangles) and transverse velocities of the H$_2$O maser features relative to the feature~\#~4 (indicated with the vertex-connected symbol). Colored cones are used to show both the direction and the uncertainty (cone aperture) of the proper motion of maser features. The proper motion amplitude scale is given by the black arrow on the bottom left corner of the panel. Different colors are used to indicate the maser LSR velocities, according to the color scale on the right-hand side of the panel, with green denoting the systemic velocity of the HMC.  The VLA 1.3~cm and 7~mm continuum emissions are given with dotted and dashed contours, respectively, using the same contour levels shown in Fig.~\ref{fig3}.
\emph{b)} absolute positions and transverse velocities of the water maser features. The plotted field of view is the same as in the upper panel and symbols and contours have the same meaning as in the upper panel. Absolute transverse velocities of water masers are affected by large uncertainties and only the mean direction (and amplitude) of motion is shown.}
\label{fig5}
\end{figure*}

\begin{landscape}
\begin{figure}
\centering
\includegraphics[angle=-90.0,scale=0.8]{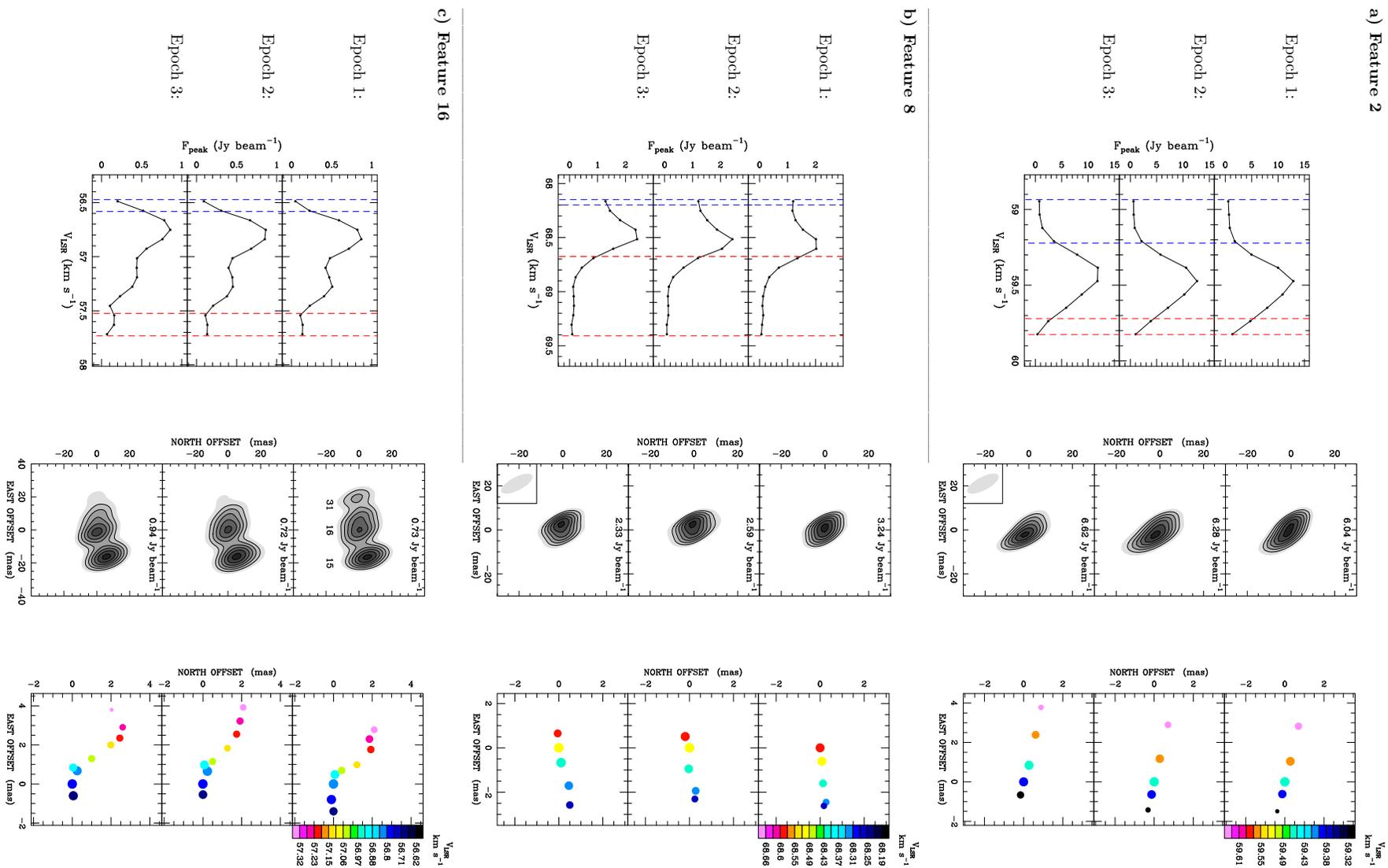}
\caption{Upper, middle, and lower set of plots present the time evolution of the spatial and spectral structure of three selected 6.7~GHz CH$_3$OH features. Left to right plots present the spectral profile, the image, and the internal velocity gradient of the feature, respectively, at our 3 observing epochs (corresponding to the upper, middle, and lower panels of each set of plots). \emph{Spectral profile:} dots report the intensities of feature's spots, emitting at different V$_{\rm LSR}$;  vertical dashed lines delimit the velocity ranges of the spectral wings, over which maser signal has been summed to produce the feature image shown in the middle plot. \emph{Image:} map of the spectral-wings emission of the feature, obtained by summing channel maps over the two velocity ranges indicated in the spectral profile.
The origin $(0, 0)$ of each map is centered on the peak position of the brightest spot of the feature. Contour levels range from 30 to $90\%$, at multiples of $10\%$, of the peak intensity reported on the top right corner of each panel.  At each epoch, the same restoring beam has been used and is shown in the lower left corner of the lower panel. The maps in the lower set of plots show the emission of three closeby, spatially blended, features, indicated with the label numbers reported in Table~\ref{met_tab}. To recover the most extended structure of this spatially-blended emission, in this case we have produced tapered maps (``UVTAPER = 10~M$\lambda$''). \emph{Internal velocity gradient:} spatial and LSR velocity distribution of the brightest spots of the feature. Different colors are used to indicate the maser LSR velocities, according to the color scale on the right-hand side of the plot. Symbol size scales logarithmically with spot intensity. In each panel, spot positions are measured from a point $(0, 0)$ which coincides with the origin of the feature image for the corresponding epoch.}
\label{fig6}
\end{figure}
\end{landscape}

\begin{figure*}[htbp]
\centering
\includegraphics[angle=-90.0,scale=0.5]{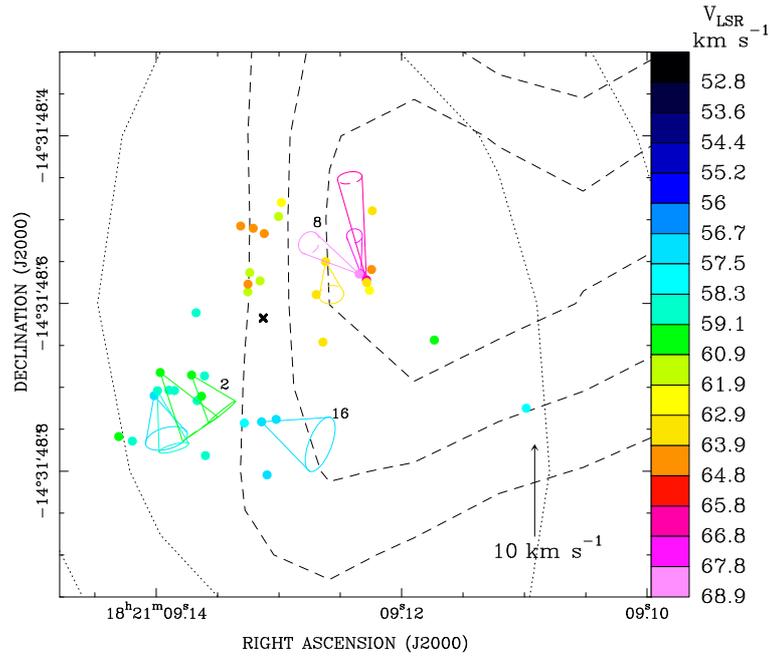}
\caption{6.7~GHz CH$_3$OH maser kinematics toward \G16. Absolute positions (dots) and transverse velocities of the CH$_3$OH maser features relative to the center of motion (as defined in Sect.~\ref{ch3oh_results}) of the methanol maser distribution (indicated by the cross). Colored cones are used to show both the direction and the uncertainty (cone aperture) of the proper motion of maser features. The proper motion amplitude scale is given by the black arrow on the bottom right corner of the panel. Different colors are used to indicate the maser LSR velocities, according to the color scale on the right-hand side of the plot, with green denoting the systemic velocity of the HMC.
The VLA 1.3~cm and 7~mm continuum emissions are given with dotted and dashed contours, respectively, using the same contour levels shown in Fig.~\ref{fig3}. Numbers close to proper-motion cones of a few features are the feature labels reported in Table~\ref{met_tab} and mark the features whose properties are presented in Fig.~\ref{fig6}.}
\label{fig7}
\end{figure*}

\begin{figure*}[htbp]
\centering
\includegraphics[angle=0.0,scale=0.7]{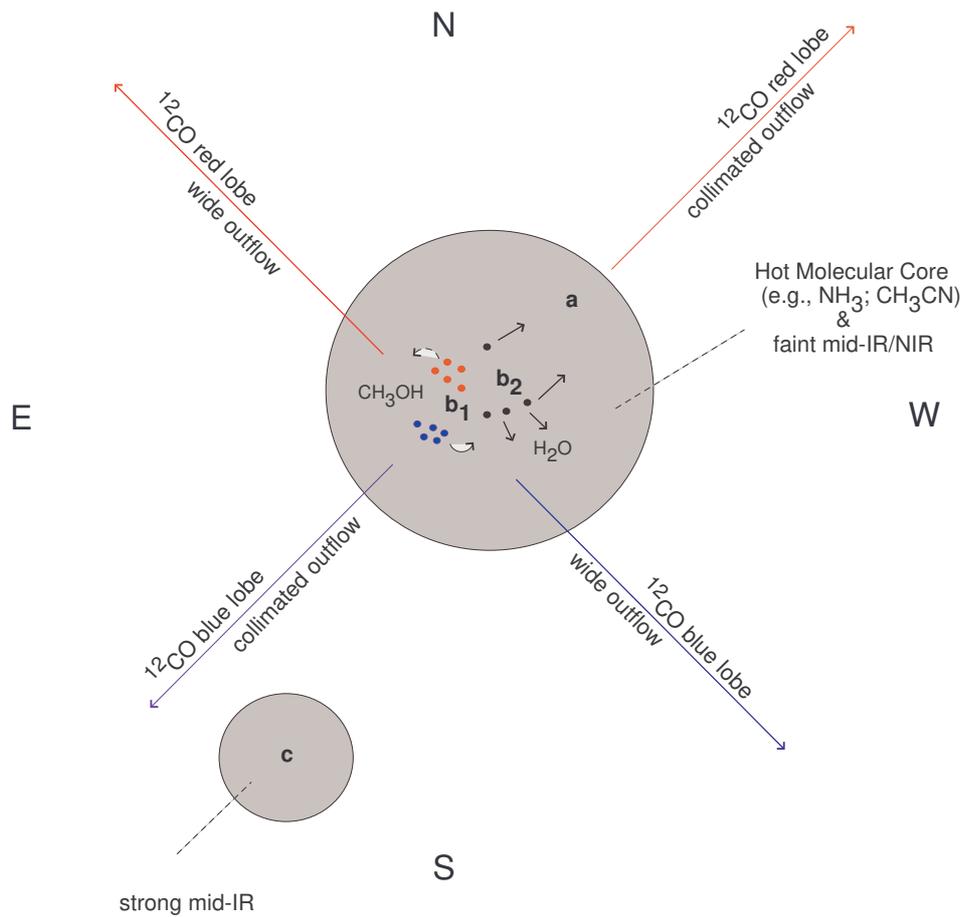}
\caption{Schematic picture of the main emission components detected toward the HMSFR \G16, within a field of view of about $10''$. The drawing is not to scale. The sources labeled ``a'', ``b'' (splitted in this paper in two components, ``b1'' and ``b2''), and ``c'' mark the three VLA radio continuum components detected by \citet{Zapata2006}. The big grey circles identify the two mid-IR sources detected by \citet{DeBuizer2005}. The northwestern one is also associated with the hot ammonia core \citep{Codella1997} and  molecular, thermal and maser, line emissions  \citep{Walsh1998,F&C1999,Beuther2006}. The main directions of the two CO bipolar outflows (NE--SW and NW--SE) resolved by \citet{Beuther2006} are shown. The distribution and main direction of motion of the methanol (blue and red dots according with approaching and receding l.o.s. velocities) and water (black dots) maser components reported in this paper are presented: 1) the CH$_3$OH masing-gas traces a disk/toroid rotating about a massive YSO, powering the NE--SW outflow and a thermal jet associated with the continuum source ``b1''; 2) the H$_2$O masing-gas together with the radio component ``b2'' and the thermal lines traces a distinct massive YSO, possibly in an earlier evolutionary phase than the one exciting the methanol masers.}
\label{fig8}
\end{figure*}

Figure~\ref{fig7} shows the 6.7~GHz CH$_3$OH masers lying on top of the 1.3~cm and 7~mm continuum source labeled ``b1'' in
Table~\ref{tab1}.
The continuum spectrum (shown in Fig.~\ref{fig2}) is consistent with the source being an
ultracompact (UC) H~\textsc{ii} region, with a turnover frequency of about 22~GHz, ionized by a B1 star. Note, however, that to fit the continuum spectrum a very small size of the UCH~\textsc{ii} region is required, of the order of a few hundredths of arcsec (red dotted line in Fig.~\ref{fig2}). That is in contrast with the
continuum emission being not detected by our 1.3~and~3.6~cm VLA--A observations (see Table~\ref{tab1}). Since in our fit we have considered an homogeneous distribution of ionized gas, one possibility is that the continuum source is {\it not} a classical  H~\textsc{ii} region, but that is characterized by strong density
and (maybe) velocity gradients.  We interpret the continuum source close to the 6.7~GHz masers in terms of free-free emission from the interaction
of a jet, powered by the YSO at the center of the 6.7~GHz maser distribution, with dense gas.
Using the fluxes measured with the VLA--C at three wavelengths (3.6, 1.3, and 0.7~cm), the derived spectral index
for the ``b1'' emission is $\alpha = 0.5$ (see Fig.~\ref{fig2}), which is consistent with the interpretation in terms of a thermal jet.

For shock induced ionization in a thermal jet, following the
calculation by \citet{Anglada1996} for optically thin emission, one has: $\rm F \, d^2 = 10^{3.5} \, (\Omega/4\pi) \, \dot{P} $, where \ F \ is the measured continuum flux in mJy, $\rm \dot{P}$ is the jet momentum rate in \mbox{M$_{\sun}$~yr$^{-1}$~\kms}, $\Omega$ is the jet solid angle in sr, and \ d \ is the source distance in kpc.  Using the flux of 1.4~mJy measured at 7~mm with the VLA--C for the source ``b1'', and the near kinematic distance of 4.4~kpc, one derives:   $\rm  \dot{P} = 8 \times 10^{-3} \, (\Omega/4\pi)^{-1}$~\mbox{M$_{\sun}$~yr$^{-1}$~\kms}.
The momentum rate thus depends on the estimate of the jet collimation factor  \ $(\Omega/4\pi)^{-1}$. Taking the center of motion of methanol masers as the YSO position, the (sky-projected) distance to the 1.3~cm continuum source (about 0\farcs1) and its size ($ \ge 1\arcsec$) imply an opening angle \ $\theta \ge  2.0 \, \arctan(0.5/0.1) = 2.7$~rad,  and a corresponding solid angle \ $\Omega \ge 5$~sr. Thus, by assuming a wide-angle wind with \ $\Omega \approx 4 \pi$,
the derived momentum rate is \ $\dot{\rm P} = 8 \times 10^{-3} $~\mbox{M$_{\sun}$~yr$^{-1}$~\kms}. From the value of the momentum measured by \citet[][Table~5]{Beuther2006}, using a dynamical time scale of  $\sim10^{4}$~yr (as derived from the outflow size of $\sim10$\arcsec\ and the average gas velocity of $\sim20$~\kms), the estimate of the momentum rate for the NE--SW $^{12}$CO~(2--1) outflow is \ $\dot{\rm P} = 6 \times 10^{-3} $~\mbox{M$_{\sun}$~yr$^{-1}$~\kms}. Such a value is consistent with that derived for the wind supposed to excite the source ``b1'', and supports our interpretation that the YSO associated with the 6.7~GHz masers can be the one driving the NE--SW $^{12}$CO~(2--1) outflow (see Fig.~\ref{fig8}).

The 7~mm VLA source ``b2'' is found offset by about 0\farcs5--0\farcs7 to the NW of the source ``b1'' (see Table~\ref{tab1} and Fig.~\ref{fig5}), in the region where emission of both 22~GHz water masers and several high-density molecular tracers are observed \citep[][Fig.~13]{Beuther2006}.
The continuum and molecular emission in this region might be powered by the same wind, emitted by the YSO associated with
the 6.7~GHz masers, which is supposed to power the source ``b1''. However, a simple evaluation of the momentum rate necessary to drive water masers, leads us to exclude this possibility. Assuming that the observed water maser emission is excited in a jet from a star, knowing the average distance of water masers from the
star and the average maser velocity, one can estimate the momentum rate of the jet (in the assumption that it is momentum-driven) with
the expression:
$$ \rm \dot{P} = 1.5 \times 10^{-3} \, V_{10}^{2} \, R_{100}^{2} \, (\Omega/4\pi)~\mbox{M$_{\sun}$~yr$^{-1}$~\kms} $$

\noindent where \ $\rm V_{10}$ \ is the average maser velocity in units of 10~\kms,  $\rm R_{100}$ \ is the average distance of water masers from the star in units of 100~AU, and \ $\Omega$ is the solid angle of the jet. In deriving this expression, following models of water maser excitation in fast shocks \citep{Elitzur1989}, we have adopted an $\rm H_{2}$ pre-shock density of \ $10^8$~cm$^{-3}$. Taking the average distance of the detected water masers from the center of motion of the 6.7~GHz masers, corresponding to \ $\approx600$~mas (or 2600~AU), the average (absolute) maser velocity of \ 60~\kms, and \ $\Omega \approx 4 \pi$ for a wide-angle wind,
the derived momentum rate is \ $ \dot{\rm P} \approx 36 $~\mbox{M$_{\sun}$~yr$^{-1}$~\kms}. This value is exceedingly too high, about four orders of magnitudes higher than that derived above from the 7~mm emission of source ``b1'', and indicates that the YSO
associated with the 6.7~GHz methanol masers is too distant from the water masers to drive their motion.

We postulate the presence of another YSO driving the motion of 22~GHz masers, powering the 7~mm continuum source
``b2'' and exciting the emission of the high-density molecular tracers. It could be possibly located close to the cluster of
strongest water masers (inset in Fig.~\ref{fig3}), which presents the largest and most scattered proper motions (see Fig.~\ref{fig5}). The ordered distribution of l.o.s. velocities with position observed for this threadlike structure of water maser features, as well as their fast variability (on a time scale between 3 to 6 months), lead us to interpret this emission in terms of a shock front driven by the nearby YSO. Both warm dust and ionized gas can contribute to the continuum source ``b2'', but having no information on the spectral index prevents us from distinguishing between the two contributions. The presence of strong water masers and the excitation of molecules as CH$_3$CN, HCOOCH$_3$, and
HNCO, typical of HMCs, suggests that this YSO can be a massive one.

The YSO exciting the 22~GHz masers is probably embedded in a denser cocoon of gas and dust than that harboring the 6.7~GHz masers.
The presence of water masers indeed witnesses high gas density with \ $\rm n_{H_{2}}  \sim 10^8$~cm$^{-3}$.
The gas density is likely to increase along the SE--NW direction till reaching the main mm
continuum peak (source ``a'' of \citealt{Zapata2006}), which marks the position of the most embedded source in the region. According to the model proposed
by \citet{Beuther2005} (see also \citealt{Arce2007}), outflows emitted by massive YSOs might increase their opening angle in the course of the YSO evolution, excavating larger
and larger cavities in the YSO natal cocoon. Following this model, more embedded massive YSOs are likely to be also less evolved.
In this view, one can speculate that the YSO at the center of the 6.7~GHz maser distribution (which we identify as the driving source of a poorly collimated $^{12}$CO~(2--1) outflow observed at larger scale) is at a later stage of evolution than that exciting the water masers, which in turn might
be more evolved than the YSO associated with the main mm continuum emission.
The most collimated
(along the SE--NW direction) $^{12}$CO~(2--1) outflow of this region, revealed by the SMA observations of \citet{Beuther2006}, could
be driven by either the YSO close to the water masers or the one responsible for the main mm emission (sketch in Fig.~\ref{fig8}).
Therefore, across a distance of $\approx0.05$~pc, there are indications of sequential star-formation
with younger and younger objects going from SE to NW.

\section{Summary and Conclusions}

Using the VLBA and the EVN interferometers, we observed the high-mass star-forming region \G16 in the three most powerful maser transitions: 22.2~GHz H$_2$O, 6.7~GHz CH$_3$CH, and 1.665~GHz OH. The radio continuum emission toward \G16 was also observed with the most extended VLA configuration to compare its brightness structure with the VLA--C archival data available. From our observations, we draw the following main conclusions:

\begin{enumerate}

\item In the present work, we have collected evidence that methanol masers, alike water emission, are suitable tracers of the kinematics around massive young stellar objects. With three VLBI epochs over a time baseline of 2~yr, we have measured accurate ($< 30\%$) proper motions of the 6.7~GHz masers. The l.o.s and sky-projected velocity distribution of methanol masers indicates that they are rotating with an average velocity of about 7~\kms\ at distances of about 600~AU from a YSO. The inferred dynamical mass (35~M$_{\odot}$) is consistent with 6.7~GHz masers tracing a high-mass YSO. The elongated NW--SE distribution of 6.7~GHz masers suggests they can originate in a flattened structure (disk/toroid) and
 that the YSO at the center of their distribution can be the one driving the motion of the NE--SW $^{12}$CO~(2--1) outflow observed at larger angular scale.

\item Close to the 6.7~GHz masers, a compact (or slightly resolved) continuum source is observed with the VLA--C at 3.6~and~1.3~cm, and at 7~mm.
We interpret this radio continuum in terms of free-free emission of gas ionized by a wide-angle wind emitted by the YSO associated
with the 6.7~GHz masers and interacting with the surrounding dense gas. The momentum rate of this  wind is consistent with that
measured for the  NE--SW $^{12}$CO~(2--1) outflow.

\item Water masers are found offset by more than 0\farcs5 from the center of the 6.7~GHz maser distribution, in the same region where
a weak 7~mm VLA source and emission in high-density molecular lines are observed. The derived absolute velocities
of the water masers, although affected by large errors, seem to indicate fast motions with an average amplitude of about 60~\kms.
We postulate the presence in this region of a distinct YSO, responsible for driving the motion of water masers and exciting  the continuum
and molecular line emissions.

\end{enumerate}

Previous SMA observations in several molecular lines have revealed multiple molecular outflows from the \G16 star-forming region, suggesting the presence of several massive YSOs across a region of $\approx0.1$~pc. This work shows that maser VLBI data are a powerful tool to identify the massive YSOs driving the large scale outflows, to study the gas kinematics close to the forming stars, and to detect disks/toroids of size of hundreds AU at typical distances of several kiloparsecs. In the next future, when VLBI maser observations will be complemented with data of new generation (sub)millimeter interferometers (i.e. ALMA) at comparable angular resolution, it will be possible to better constrain the physical and kinematical properties of both outflow(s) and core(s) hosting maser activity.

\begin{acknowledgements}

This work is partially supported by a Grant-in-Aid from the Ministry of Education,
Culture, Sports, Science and Technology of Japan (No. 20740113).

\end{acknowledgements}

\nocite{*}
\bibliographystyle{aa}
\bibliography{sanna_G16}

\clearpage

\clearpage

\addtocounter{table}{0}

\begin{table*}
\caption{G16.59--0.05: Radio continuum associated with the CH$_3$OH and H$_2$O maser emission. \label{tab3}}
\label{tab1}
\begin{tabular}{c l c c c c c c c c}

\hline \hline
        &          &            &              &           &                   & \multicolumn{2}{c}{ Peak position } &         &            \\
  Label &Telescope & $\lambda$  &    HPBW      &    P.A.   &    Image rms      & R.A.(J2000) &  Dec.(J2000) &   F$_{\rm peak}$ &  F$_{\rm int}$ \\
        &          &    (cm)    & ($'' \times ''$) &  (\degr)  & (mJy beam$^{-1}$) &   (h m s)   &  (\degr $'$ $''$)& (mJy beam$^{-1}$)&   (mJy)    \\

\hline
& & & & & & & & & \\
b1 & VLA--C$^1$     &    0.7     &  $ 0.9   \times  0.5  $  &    -10    &  0.2  &  18 21 09.12   &  -14 31 48.5  &   0.7   & 1.4 \\
   & VLA--A     &    1.3     &  $ 0.142 \times  0.096$  &    12    &  0.09 &    ...         &          ...  &   $<0.27(3\sigma)$ & ... \\
   & VLA--C$^1$     &    1.3     &  $ 1.7   \times  1.0  $  &    22    &  0.07 &  18 21 09.12   &  -14 31 48.7  &   0.6   & 1.3 \\
   & VLA--A     &    3.6     &  $0.39   \times  0.24 $  &   -5     &  0.03 &    ...         &     ...       &  $<0.09(3\sigma)$  & ... \\
   & VLA--C     &    3.6     &  $4.2    \times  2.6  $  &    2     &  0.03 &  18 21 09.11   &  -14 31 48.5  &   0.37  & 0.61 \\
   & VLA--B$^2$     &    6.0     &  $ 1.5   \times  1.5  $  &    0    &  0.3  &    ...         &     ...       &   $<0.9(3\sigma)$ & ... \\
b2 & VLA--C$^1$     &    0.7     &  $ 0.9   \times  0.5  $  &    -10    &  0.2  &  18 21 09.08   &  -14 31 48.3 &    0.7   & 1.1 \\
& & & & & & & & & \\
& & & & & & & & & \\
\hline
& & & & & & & & & \\
\end{tabular}


\footnotesize{(1) \
Observations from archival data, project code AZ149 (P.I. Zapata).}

\footnotesize{(2) \
Data from the CORNISH survey: ``The CORNISH Survey of the Galactic Plane", C. R. Purcell, M. G. Hoare, P. Diamond, 2008, Astronomical Society of the Pacific Conference Series, Vol. 387, Pg. 389.}


\end{table*}

\clearpage

\clearpage

\clearpage

\clearpage

\clearpage

\clearpage

\clearpage

\onllongtab{2}{
\begin{longtable}{ccrcrrcrrr}

\caption{\emph{Parameters of VLBA 22.2~GHz water maser features.}
For each identified feature, the label (given in Col.~1) increases with decreasing brightness.
Cols.~2~and~3 report the LSR velocity and brightness of the brightest spot (for the epoch specified by the number in brackets).
Cols.~4~and~5 give the position offset relative to the feature~\#~4, toward the East and the North directions, respectively.
Cols.~6~and~7 report the projected components of the feature proper motion relative the the feature~\#~4, along the East and North directions, respectively. Col.~8 shows the relative error of the proper motion amplitude.
The absolute position of the reference feature~\#~4 is reported at the bottom of the Table.} \label{wat_tab} \\

\hline \hline
Feature &  $V_{\rm LSR}$ & \multicolumn{1}{c}{F$_{\rm peak}$}  & & \multicolumn{1}{c}{$\Delta x$} & \multicolumn{1}{c}{$\Delta y$} &  & \multicolumn{1}{c}{$V_{\rm x}$} & \multicolumn{1}{c}{$V_{\rm y}$} & \multicolumn{1}{c}{$\Delta|V|/|V| $} \\
        &  (km s$^{-1}$) & \multicolumn{1}{c}{(Jy beam$^{-1}$)} &  & \multicolumn{1}{c}{(mas)} & \multicolumn{1}{c}{(mas)} &  & \multicolumn{1}{c}{(km s$^{-1}$)} & \multicolumn{1}{c}{(km s$^{-1}$)} &                  \\
\hline
& & & & & & & & &\\
\endfirsthead
\caption{continued.}\\
\hline \hline
Feature &  $V_{\rm LSR}$ & \multicolumn{1}{c}{F$_{\rm peak}$}  & & \multicolumn{1}{c}{$\Delta x$} & \multicolumn{1}{c}{$\Delta y$} &  & \multicolumn{1}{c}{$V_{\rm x}$} & \multicolumn{1}{c}{$V_{\rm y}$} & \multicolumn{1}{c}{$\Delta|V|/|V| $} \\
        &  (km s$^{-1}$) & \multicolumn{1}{c}{(Jy beam$^{-1}$m)} &  & \multicolumn{1}{c}{(mas)} & \multicolumn{1}{c}{(mas)} &  & \multicolumn{1}{c}{(km s$^{-1}$)} & \multicolumn{1}{c}{(km s$^{-1}$)} &                  \\
\hline
& & & & & & & & &\\
\endhead
& & & & & & & & &\\
\hline
\endfoot
& & & & & & & & &\\
\hline
\endlastfoot
1    & 65.48 & 65.49 (2)& & $ -106.46\pm 0.23 $ & $ 132.43\pm0.44 $ & & $ -17.4\pm6.9 $ & $  40.8\pm12.9$ & 27\% \\
2    & 64.85 & 40.19 (4)& & $ -110.94\pm 0.18 $ & $ 142.62\pm0.18 $ & &      ...        &      ...        & ...  \\
3    & 65.06 & 27.55 (3)& & $ -108.89\pm 0.12 $ & $ 139.37\pm0.17 $ & & $ -23.3\pm5.5 $ & $  42.8\pm7.8 $ & 15\% \\
4    & 60.93 & 22.98 (1)& & $          0      $ & $          0    $ & &       0        &        0         &      \\
5    & 60.09 & 15.89 (1)& & $  107.14\pm 0.10 $ & $ 109.94\pm0.19 $ & & $  -0.1\pm7.0 $ & $  39.7\pm13.3$ & 33\% \\
6    & 64.52 & 11.63 (1)& & $ -111.92\pm 0.12 $ & $ 143.40\pm0.11 $ & & $  -3.3\pm4.0 $ & $  19.7\pm4.4 $ & 22\% \\
7    & 64.42 &  4.24 (4)& & $ -107.50\pm 0.09 $ & $ 131.83\pm0.11 $ & & $ -10.6\pm5.7 $ & $  43.1\pm6.6 $ & 15\% \\
8    & 61.48 &  3.30 (4)& & $  230.00\pm 0.10 $ & $ 685.88\pm0.18 $ & & $  -7.2\pm7.0 $ & $  35.4\pm11.7$ & 32\% \\
9    & 51.36 &  2.70 (4)& & $  256.26\pm 0.16 $ & $ 695.78\pm0.17 $ & & $  -1.8\pm8.5 $ & $  55.3\pm8.1 $ & 15\% \\
10   & 59.25 &  2.56 (1)& & $  229.96\pm 0.13 $ & $ 685.76\pm0.20 $ & &      ...        &      ...        & ...  \\
11   & 62.62 &  1.72 (1)& & $ -111.69\pm 0.09 $ & $ 120.16\pm0.10 $ & &      ...        &      ...        & ...  \\
12   & 64.42 &  1.70 (4)& & $ -105.56\pm 0.08 $ & $ 137.21\pm0.10 $ & &      ...        &      ...        & ...  \\
13   & 66.11 &  1.67 (3)& & $ -116.31\pm 0.09 $ & $ 156.20\pm0.10 $ & & $ -10.9\pm6.0 $ & $  43.5\pm8.5 $ & 19\% \\
14   & 60.84 &  1.48 (3)& & $  235.32\pm 0.08 $ & $ 673.64\pm0.10 $ & & $ -11.8\pm5.4 $ & $  19.7\pm6.1 $ & 26\% \\
15   & 61.26 &  1.31 (2)& & $  107.55\pm 0.07 $ & $ 111.31\pm0.09 $ & &      ...        &      ...        & ...  \\
16   & 60.63 &  1.11 (3)& & $    0.56\pm 0.14 $ & $  -1.27\pm0.13 $ & &      ...        &      ...        & ...  \\
17   & 66.74 &  0.87 (4)& & $ -115.99\pm 0.09 $ & $ 153.86\pm0.21 $ & &      ...        &      ...        & ...  \\
18   & 65.06 &  0.80 (3)& & $ -118.02\pm 0.11 $ & $ 153.03\pm0.18 $ & &      ...        &      ...        & ...  \\
19   & 61.90 &  0.79 (4)& & $ -108.63\pm 0.13 $ & $ 131.99\pm0.17 $ & &      ...        &      ...        & ...  \\
20   & 66.42 &  0.72 (1)& & $ -106.92\pm 0.10 $ & $ 129.23\pm0.18 $ & &      ...        &      ...        & ...  \\
21   & 65.48 &  0.71 (4)& & $  106.92\pm 0.08 $ & $ 110.26\pm0.09 $ & &      ...        &      ...        & ...  \\
22   & 59.58 &  0.64 (4)& & $  152.99\pm 0.09 $ & $   1.63\pm0.10 $ & & $   9.3\pm3.5 $ & $  11.7\pm4.2 $ & 27\% \\
23   & 67.89 &  0.57 (1)& & $ -115.27\pm 0.09 $ & $ 146.15\pm0.09 $ & &      ...        &      ...        & ...  \\
24   & 62.53 &  0.51 (4)& & $ -112.30\pm 0.08 $ & $ 126.55\pm0.09 $ & &      ...        &      ...        & ...  \\
25   & 66.74 &  0.49 (2)& & $ -115.84\pm 0.08 $ & $ 157.32\pm0.12 $ & & $ -27.4\pm5.0 $ & $ 113.8\pm7.6 $ &  6\% \\
26   & 64.10 &  0.46 (1)& & $ -106.47\pm 0.10 $ & $ 137.54\pm0.12 $ & &      ...        &      ...        & ...  \\
27   & 66.32 &  0.45 (3)& & $ -117.63\pm 0.08 $ & $ 162.42\pm0.12 $ & &      ...        &      ...        & ...  \\
28   & 65.48 &  0.40 (2)& & $  107.06\pm 0.09 $ & $ 109.40\pm0.10 $ & &      ...        &      ...        & ...  \\
29   & 62.74 &  0.39 (3)& & $ -112.00\pm 0.09 $ & $ 118.35\pm0.14 $ & &      ...        &      ...        & ...  \\
30   & 60.30 &  0.35 (1)& & $    6.92\pm 0.09 $ & $  12.23\pm0.12 $ & & $   4.9\pm3.6 $ & $  20.4\pm4.6 $ & 21\% \\
31   & 60.63 &  0.34 (4)& & $  107.16\pm 0.11 $ & $ 111.54\pm0.18 $ & &      ...        &      ...        & ...  \\
32   & 62.20 &  0.33 (1)& & $ -109.48\pm 0.19 $ & $ 128.05\pm0.20 $ & &      ...        &      ...        & ...  \\
33   & 60.84 &  0.25 (4)& & $    5.81\pm 0.09 $ & $  10.10\pm0.14 $ & &      ...        &      ...        & ...  \\
34   & 60.84 &  0.21 (3)& & $   -8.41\pm 0.09 $ & $  13.03\pm0.20 $ & &      ...        &      ...        & ...  \\
35   & 62.32 &  0.21 (4)& & $ -111.81\pm 0.08 $ & $ 120.87\pm0.12 $ & &      ...        &      ...        & ...  \\
36   & 63.46 &  0.18 (1)& & $ -107.52\pm 0.10 $ & $ 131.67\pm0.13 $ & &      ...        &      ...        & ...  \\
37   & 60.09 &  0.16 (1)& & $  106.91\pm 0.10 $ & $ 119.47\pm0.15 $ & &      ...        &      ...        & ...  \\
38   & 58.61 &  0.10 (1)& & $  228.34\pm 0.10 $ & $ -31.43\pm0.18 $ & & $  12.8\pm4.3 $ & $  10.4\pm7.0 $ & 34\% \\
39   & 63.04 &  0.10 (1)& & $  264.41\pm 0.11 $ & $ 684.83\pm0.15 $ & &      ...        &      ...        & ...  \\
40   & 60.00 &  0.10 (3)& & $  108.53\pm 0.09 $ & $ 113.03\pm0.15 $ & &      ...        &      ...        & ...  \\

& & & & & & & & &\\
& & & & & & & & &\\
\multicolumn{10}{c}{Reference Feature: Absolute position}\\
\hline\hline
Feature & \multicolumn{4}{c}{R.A.(J2000)} & \multicolumn{4}{c}{ Dec.(J2000) }   &   \\
        & \multicolumn{4}{c}{(h m s)}     & \multicolumn{4}{c}{(\degr $'$ $''$)}    &   \\
\hline
& & & & & & & & &\\
4 &\multicolumn{4}{c}{18:21:09.09211$\pm$0.00009}&\multicolumn{4}{c}{-14:31.48.6790$\pm$0.0023} & \\

\end{longtable}
}

\onllongtab{3}{
\begin{longtable}{ccrrcrrcrrr}

\caption{\emph{Parameters of EVN 6.7~GHz methanol maser features.}
For each identified feature, the label (given in Col.~1) increases with decreasing brightness.
Cols.~2,~3~and~4 report the LSR velocity and brightness of the brightest spot, and its percent brightness
variability.
Cols.~5~and~6 give the position offset relative to the feature~\#~2, toward the East and the North directions, respectively.
Cols.~7~and~8 report the projected components of the feature proper motion relative to the center of motion (as defined in Sect.~\ref{ch3oh_results}, and identified with label~\#~0), along the East and North directions, respectively.
Col.~9 shows the relative error of the proper motion amplitude.
The absolute position of the reference feature~\#~2 is reported at the bottom of the Table.} \label{met_tab} \\

\hline
\hline
Feature &  $V_{\rm LSR}$ & \multicolumn{1}{c}{F$_{\rm peak}$} & Var.$^a$ & & \multicolumn{1}{c}{$\Delta x$} & \multicolumn{1}{c}{$\Delta y$} &  & \multicolumn{1}{c}{$V_{\rm x}$} & \multicolumn{1}{c}{$V_{\rm y}$} & \multicolumn{1}{c}{$\Delta|V|/|V| $} \\
        &  (km s$^{-1}$) & \multicolumn{1}{c}{(Jy beam$^{-1}$)} & &  & \multicolumn{1}{c}{(mas)} & \multicolumn{1}{c}{(mas)} &  & \multicolumn{1}{c}{(km s$^{-1}$)} & \multicolumn{1}{c}{(km s$^{-1}$)} &                  \\
\hline
 & & & & & & & & & &\\
\endfirsthead
\caption{continued.}\\
\hline \hline
Feature &  $V_{\rm LSR}$ & \multicolumn{1}{c}{F$_{\rm peak}$} & Var.$^a$ & & \multicolumn{1}{c}{$\Delta x$} & \multicolumn{1}{c}{$\Delta y$} &  & \multicolumn{1}{c}{$V_{\rm x}$} & \multicolumn{1}{c}{$V_{\rm y}$} & \multicolumn{1}{c}{$\Delta|V|/|V| $} \\
        &  (km s$^{-1}$) & \multicolumn{1}{c}{(Jy beam$^{-1}$)} & &  & \multicolumn{1}{c}{(mas)} & \multicolumn{1}{c}{(mas)} &  & \multicolumn{1}{c}{(km s$^{-1}$)} & \multicolumn{1}{c}{(km s$^{-1}$)} &                  \\
\hline
 & & & & & & & & & &\\
\endhead
 & & & & & & & & & &\\
\hline
\endfoot
 & & & & & & & & & &\\
\hline
\endlastfoot
1   & 62.11 & 20.62 & 20\% & & $ -106.27\pm0.09 $ & $ 205.81\pm0.13 $ & &       ...        &       ...        & ...  \\
2   & 59.47 & 12.90 & 8\%  & & $           0    $ & $         0     $ & & $ -3.2 \pm 1.0 $ & $ -4.0 \pm 1.4 $ & 25\% \\
3   & 64.13 &  7.88 & 20\% & & $  -72.93\pm0.09 $ & $ 175.07\pm0.13 $ & &       ...        &       ...        & ...  \\
4   & 58.42 &  7.60 & 3\%  & & $   40.35\pm0.09 $ & $ -18.67\pm0.13 $ & & $ -1.7 \pm 1.0 $ & $ -6.1 \pm 1.4 $ & 22\% \\
5   & 63.78 &  7.15 & 35\% & & $ -158.52\pm0.09 $ & $ 135.34\pm0.13 $ & & $ -0.6 \pm 1.0 $ & $ -3.5 \pm 1.4 $ & 40\% \\
6   & 66.06 &  4.15 & 9\%  & & $ -206.96\pm0.09 $ & $ 113.62\pm0.13 $ & & $  1.8 \pm 1.1 $ & $ 11.0 \pm 1.5 $ & 14\% \\
7   & 67.81 &  2.98 & 18\% & & $ -200.44\pm0.08 $ & $ 120.97\pm0.12 $ & & $  0.7 \pm 1.0 $ & $  4.1 \pm 1.4 $ & 35\% \\
8   & 68.52 &  2.42 & 18\% & & $ -198.32\pm0.10 $ & $ 120.56\pm0.14 $ & & $  5.4 \pm 1.2 $ & $  3.3 \pm 1.7 $ & 22\% \\
9   & 63.95 &  2.03 & 17\% & & $  -58.07\pm0.10 $ & $ 177.76\pm0.17 $ & &       ...        &       ...        & ...  \\
10  & 59.91 &  1.77 & 45\% & & $   37.02\pm0.10 $ & $   3.00\pm0.15 $ & & $ -4.2 \pm 1.2 $ & $ -6.1 \pm 1.8 $ & 22\% \\
11  & 64.74 &  1.60 & 32\% & & $ -212.91\pm0.10 $ & $ 125.86\pm0.16 $ & &       ...        &       ...        & ...  \\
12  & 57.02 &  1.20 & 21\% & & $   44.19\pm0.09 $ & $ -24.65\pm0.16 $ & & $ -1.3 \pm 1.2 $ & $ -4.6 \pm 2.0 $ & 41\% \\
13  & 58.42 &  1.16 & 8\%  & & $   20.18\pm0.18 $ & $ -18.57\pm0.46 $ & &       ...        &       ...        & ...  \\
14  & 64.30 &  1.13 & 30\% & & $  -85.95\pm0.10 $ & $ 168.74\pm0.17 $ & &       ...        &       ...        & ...  \\
15  & 57.45 &  0.99 & 10\% & & $ -100.07\pm0.10 $ & $ -52.87\pm0.16 $ & &       ...        &       ...        & ...  \\
16  & 56.84 &  0.87 & 2\%  & & $  -82.83\pm0.11 $ & $ -55.85\pm0.18 $ & & $ -6.3 \pm 1.6 $ & $ -2.4 \pm 2.6 $ & 27\% \\
17  & 62.55 &  0.72 & 44\% & & $ -210.41\pm0.13 $ & $ 101.17\pm0.27 $ & &       ...        &       ...        & ...  \\
18  & 58.51 &  0.56 &...   & & $  -15.95\pm0.22 $ & $   0.44\pm0.33 $ & &       ...        &       ...        & ...  \\
19  & 58.60 &  0.49 &...   & & $   69.45\pm0.19 $ & $ -77.32\pm0.31 $ & &       ...        &       ...        & ...  \\
20  & 61.84 &  0.41 & 25\% & & $ -102.92\pm0.15 $ & $ 189.05\pm0.30 $ & &       ...        &       ...        & ...  \\
21  & 61.76 &  0.40 & 45\% & & $  -81.04\pm0.19 $ & $ 112.36\pm0.46 $ & &       ...        &       ...        & ...  \\
22  & 58.86 &  0.28 & 34\% & & $   -6.48\pm0.18 $ & $ -30.55\pm0.39 $ & &       ...        &       ...        & ...  \\
23  & 64.04 &  0.18 &...   & & $  -68.46\pm0.42 $ & $ 103.02\pm0.58 $ & &       ...        &       ...        & ...  \\
24  & 59.12 &  0.17 &...   & & $   84.43\pm0.39 $ & $ -78.86\pm0.66 $ & &       ...        &       ...        & ...  \\
25  & 60.09 &  0.17 & 10\% & & $  -11.76\pm0.22 $ & $ -25.28\pm0.50 $ & &       ...        &       ...        & ...  \\
26  & 58.68 &  0.16 &...   & & $  -17.03\pm0.63 $ & $ -94.58\pm0.91 $ & &       ...        &       ...        & ...  \\
27  & 59.12 &  0.16 & 17\% & & $   26.70\pm0.24 $ & $ -18.15\pm0.63 $ & &       ...        &       ...        & ...  \\
28  & 58.07 &  0.14 & 19\% & & $ -396.02\pm0.25 $ & $ -39.64\pm0.64 $ & &       ...        &       ...        & ...  \\
29  & 63.07 &  0.13 & 44\% & & $ -213.83\pm0.28 $ & $ 195.95\pm0.63 $ & &       ...        &       ...        & ...  \\
30  & 57.63 &  0.12 &...   & & $  -62.37\pm0.30 $ & $ -57.29\pm0.58 $ & &       ...        &       ...        & ...  \\
31  & 61.76 &  0.12 &...   & & $  -68.25\pm0.42 $ & $  93.82\pm0.77 $ & &       ...        &       ...        & ...  \\
32  & 60.70 &  0.11 &...   & & $ -287.88\pm0.56 $ & $ -43.17\pm0.80 $ & &       ...        &       ...        & ...  \\
33  & 58.77 &  0.09 &...   & & $   -7.14\pm0.97 $ & $  68.86\pm1.10 $ & &       ...        &       ...        & ...  \\
34  & 61.32 &  0.09 & 15\% & & $  -68.81\pm0.33 $ & $ 122.23\pm0.80 $ & &       ...        &       ...        & ...  \\
35  & 63.25 &  0.08 &...   & & $ -157.22\pm0.65 $ & $  33.89\pm1.01 $ & &       ...        &       ...        & ...  \\
36  & 63.51 &  0.08 &...   & & $ -209.00\pm0.54 $ & $ 104.94\pm1.03 $ & &       ...        &       ...        & ...  \\
37  & 63.69 &  0.08 &...   & & $ -149.29\pm0.55 $ & $  90.60\pm0.57 $ & &       ...        &       ...        & ...  \\
38  & 52.89 &  0.07 &...   & & $ -407.11\pm0.65 $ & $ 669.44\pm1.07 $ & &       ...        &       ...        & ...  \\
39  & 56.84 &  0.06 &...   & & $  -91.01\pm0.95 $ & $-124.54\pm1.31 $ & &       ...        &       ...        & ...  \\

 & & & & & & & & & &\\
0     &        &       &      & & $ -84.71 \pm 0.06$ & $ 67.90 \pm 0.10$ & &  $   0  $  &   $   0  $  & \\

 & & & & & & & & & &\\
 & & & & & & & & & &\\
\multicolumn{10}{c}{Reference Feature: Absolute position}\\
\hline\hline
Feature & \multicolumn{4}{c}{R.A.(J2000)} & \multicolumn{4}{c}{ Dec.(J2000) } & & \\
        & \multicolumn{4}{c}{(h m s)}     & \multicolumn{4}{c}{(\degr $'$ $''$)}  & & \\
\hline
 & & & & & & & & & &\\
2 &\multicolumn{4}{c}{18:21:09.13711$\pm$0.00023}&\multicolumn{4}{c}{-14:31:48.6853$\pm$0.0036} & & \\

\end{longtable}

\footnotesize{(a) \
Variability (\%) = (F$_{\rm MAX}$ $-$ F$_{\rm MIN}$ of the brightest spot) / ((F$_{\rm MAX}$ + F$_{\rm MIN}$)/2.0)\\  }
}

\end{document}